\documentclass{aa}  

\usepackage{pifont}
\usepackage{amsmath}
\usepackage{graphicx}
\usepackage{txfonts}
\usepackage{url}
\usepackage{longtable}
\usepackage{lscape}
\usepackage{natbib}
\usepackage[colorlinks,linkcolor=red,anchorcolor=green,citecolor=blue]{hyperref}

\newcommand{\HII}{H\,{\small{II}} }
\newcommand\hii{\mbox{H \small{II}\normalsize}}

\begin{document}

   \title{The effects of ionization feedback on star formation: \\  A case study of the M16  H II region }
  \author{Jin-Long Xu\inst{1,6},
          Annie Zavagno\inst{2},
          Naiping Yu\inst{1},
          Xiao-Lan Liu\inst{1},
           Ye Xu\inst{3,7},
           Jinghua Yuan\inst{1},
           Chuan-Peng Zhang\inst{1},
           Si-Ju Zhang\inst{2},
           Guo-Yin Zhang\inst{1},
           Chang-Chun Ning\inst{4},
          \ and Bing-Gang Ju\inst{5,7}
          }
   \institute{National Astronomical Observatories, Chinese Academy of Sciences,
             Beijing 100101, China \\
         \email{xujl@bao.ac.cn}
         \and
         Aix Marseille Univ, CNRS, LAM, Laboratoire d'Astrophysique de Marseille, Marseille, France
          \and
          Purple Mountain Observatory, Chinese Academy of Sciences, Nanjing 210008, China
          \and
          Tibet University, Lhasa,  Tibet 850000, China
          \and
          Purple Mountain Observatory, Qinghai Station,Delingha 817000, China
         \and
      CAS Key Laboratory of FAST, National Astronomical Observatories, Chinese Academy of Sciences, Beijing 100101, China
          \and Key Laboratory of Radio Astronomy, Chinese Academy of Sciences, China
        }
\authorrunning{J.-L. Xu et. al.}
\titlerunning{The effects of ionization feedback on star formation}

   \date{Received XXX, XXX; accepted XXX, XXX}

\abstract
   {}
{We aim to investigate the impact of the ionized radiation from the M16 \HII region on the surrounding molecular cloud and on its hosted star formation.}
{To present  comprehensive multi-wavelength observations towards the M16 \HII region, we used new CO data and existing infrared, optical, and submillimeter data. The $^{12}$CO $J$=1-0, $^{13}$CO $J$=1-0, and C$^{18}$O $J$=1-0 data were obtained  with the Purple Mountain Observatory (PMO) 13.7m radio telescope. To trace massive clumps and extract young stellar objects (YSOs) associated with the M16 \HII region, we used the ATLASGAL and GLIMPSE I catalogs, respectively.}
{From CO data, we  discern a large-scale  filament with three velocity components. Because these three components overlap with each other in both velocity and space, the filament may be made of three layers.  The M16 ionized gas interacts with the large-scale filament and has reshaped its structure. In the  large-scale filament, we find 51 compact cores from the ATLASGAL catalog, 20 of them being  quiescent.  The mean excitation temperature of these cores is 22.5 K, while this is 22.2 K for the quiescent cores. This high temperature observed for the quiescent cores suggests that the cores may be heated by M16 and do not experience internal heating from sources in the cores. Through the relationship between the mass and radius of these cores, we obtain that 45\% of all the cores are massive enough to potentially form massive stars. Compared with the thermal motion, the turbulence created by the nonthermal motion is responsible for the core formation. For the pillars observed towards M16, the H II region may give rise to the strong turbulence. }
   {}


   \maketitle
%

\section{Introduction}
Filamentary infrared dark clouds (IRDC) are regarded as the precursors of massive stars and star clusters  \citep{Egan1998,Carey2000,Rathborne2006}.  \citet{Jackson2010} outline the evolutive processes of a filamentary IRDC. When a massive star forms inside a filament,  its ultraviolet (UV) radiation will ionize and heat the surrounding gas to create an \HII region. The expanding \HII region will reshape the surrounding molecular gas of the filament and even directly trigger star formation by collect-and-collapse and/or radiative driven implosion processes \citep[e.g.,][]{Elmegreen1977, Deharveng2003, Zavagno2006, Zavagno2007, Anderson2012, Dewangan2013, Xu2014, Samal2015, Xu2016}. On the other hand, in a filament, the formation of cores and clumps may be regulated by the interplay between gravity, turbulence, and magnetic field \citep{Li2015}. However, the turbulence has been shown to dissipate quickly \citep{Stone1998,Mac1999} if the external driving is stopped, resulting in the need to continuously drive turbulence via stellar feedback \citep{Ostriker2010,Offner2018}. \HII regions containing kinetic, ionization, and thermal energies can help to maintain the observed turbulence in a filamentary molecular cloud by means of injection of these energies \citep{Narayanan2008,Arce2011,Xu2018}. However, for an \HII region, the ionization energy is one order of magnitude higher than the kinetic energy and thermal energy \citep{Freyer2003,Xu2018}.  In the G47.06+0.26 filament, \citet{Xu2018} obtained that the ionization energy from  \HII regions is roughly equal to the turbulent energy, and it may help to maintain the observed turbulence.  Hence, feedback by ionizing radiation provides one possible way to regulate star formation. This regulation is positive (increase in star formation) or negative (decrease in star formation), as shown by numerical simulations for the ionizing radiation \citep[e.g.,][]{Geen2017,Kim2017,Haid2018}.  Although \HII regions may play a significant role in the self-regulation of star formation \citep[e.g.,][]{Norman1980,Dewangan2016,Dewangan2018,Xu2017}, it remains unclear which mechanisms from the feedback of a \HII region dominate for  cores, clumps, and even stars formation in a filament.

The Eagle Nebula, M16, is an optically visible Galactic \HII region located at a distance of $\sim$2 kpc \citep{Hillenbrand1993}. This \HII region is ionized by numerous O and B stars within an open cluster NGC6611 \citep{Hillenbrand1993}. The age of the young stellar objects (YSOs) associated with NGC6611 is estimated to be 0.25--3 Myr \citep{Hillenbrand1993}.  An H I shell with the velocity component in the 20--30 km s$^{-1}$ range is observed surrounding M16 \citep{Sofue1986}. Using the Columbia $^{12}$CO-line survey, \citet{Handa1986}  detected a giant molecular cloud (GMC) in contact with the northeastern edge of the shell. Hence, the expansion of the shell may be blocked by this GMC in the northeastern side.  Using the {\it Herschel} data, \citet{Hill2012} revealed several filaments in the GMC. The M16 \HII region  is also well known  because the surrounding gas exhibits a number of spectacular pillars, with the  heads faced towards the central part of NGC6611 and the tails trailed towards the opposite side.  Three of these pillars became famous as the Pillars of Creation \citep{Hester1996}. \citet{Sugitani2002}  found that a number of YSO candidates are associated with these head--tail pillars, while \citet{Andersen2004} also identified some cores in one of the pillars and demonstrated that star formation is taking place in this pillar. Hence, M16 is also known to be an active star-forming region.
 
In this paper, we present new maps of $^{12}$CO $J$=1-0, $^{13}$CO $J$=1-0, and C$^{18}$O $J$=1-0 obtained  with the Purple Mountain Observatory (PMO) 13.7m radio telescope. Combining our data with existing infrared and submillimeter images and spectra, described in Sect. 2.2, we aim to study the impact of the ionized radiation from  the M16 \HII region on the surrounding molecular cloud and on its hosted star formation. Our observations and data reduction are described in Sect. 2, and the results are presented in Sect. 3. In Sect. 4, we  discuss gas structure associated with M16, core properties and formation, and  turbulent fragmentation in a clump and a southern filament, while our conclusions are summarized in Sect. 5.  

\section{Observation and data processing}
\label{sect:data}
\subsection{CO-Lines Data}
We map a $45^{\prime}\times 45^{\prime}$ region centered at l=17.023$^{\circ}$, b=0.871$^{\circ}$ in the transitions of $^{12}$CO $J$=1-0 (115.271 GHz), $^{13}$CO $J$=1-0 (110.201 GHz), and C$^{18}$O $J$=1-0 (109.782 GHz) lines using the PMO 13.7m radio telescope at De Ling Ha in western China, during April 2018. The 3$\times$3 beam array receiver system in single-sideband (SSB) mode was used as front end. The back end is a Fast Fourier Transform Spectrometer (FFTS) of 16384 channels with a bandwidth of 1 GHz, corresponding to a velocity resolution of 0.16 km s$^{-1}$ for $^{12}$CO $J$=1-0, and 0.17 km s$^{-1}$ for $^{13}$CO $J$=1-0 and C$^{18}$O $J$=1-0. $^{12}$CO $J$=1-0 was observed at upper sideband with a system noise temperature (Tsys) of 320 K, while $^{13}$CO $J$=1-0 and C$^{18}$O $J$=1-0 were observed simultaneously at lower sideband with a  system noise temperature of 162 K.  The half-power beam width (HPBW) was 53$^{\prime\prime}$ at 115 GHz and the main beam efficiency was 0.5. The pointing accuracy of the telescope was better than 5$^{\prime\prime}$. Mapping observations  use the on-the-fly mode with a constant integration time of 14 seconds at each point and with a $0.5^{\prime}\times0.5^{\prime}$ grid.  The final data were recorded in brightness temperature scale of $T_{\rm mb}$ (K) and were reduced using the GILDAS/CLASS\footnote{http://www.iram.fr/IRAMFR/GILDAS/} package.

\subsection{Archival data}
\label{sect:archive}
In order to obtain a complete view of the filament around M16, we completed our observations with  Galactic Ring Survey (GRS)  data of $^{13}$CO $J$=1-0 emission \citep{Jackson2006}. The survey covers a longitude range of $\ell$$=$18$^{\circ}$--55.7$^{\circ}$ and a latitude range of $|b|$$\leq$1$^{\circ}$, with an  angular resolution of 46$^{\prime\prime}$.  At the velocity resolution of 0.21 km s$^{-1}$, the typical rms sensitivity is 0.13 K.

The GLIMPSE survey observed the Galactic plane (65$^{\circ} < |l| <  10^{\circ}$ for $|b| < 1^{\circ}$) with the four IR bands (3.6, 4.5, 5.8, and 8.0 $\mu$m) of the Infrared Array Camera (IRAC) \citep{Benjamin2003} on the {\it Spitzer} Space Telescope. We used these data to identify  young stars and H {\small II} regions. The spatial resolution in the four IR bands are from 1.5$^{\prime\prime}$ to 1.9$^{\prime\prime}$.

To explore the dust distribution, we used the 250 $\mu$m data from the {\it {\it Herschel}} Infrared Galactic Plane survey \citep[Hi-GAL;][]{Molinari2010} carried out by the {\it {\it Herschel}} Space Observatory. The initial survey covered a Galactic longitude region of 300$^{\circ}$$<$$\ell$$<$60$^{\circ}$ and $|b|$$<$1.0$^{\circ}$.  The angular resolution of the 250 $\mu$m band is 18$\farcs2$.  Additionally, we extracted 870 $\mu$m data from the ATLASGAL survey \citep{Schuller2009}. The survey was carried out with the Large APEX Bolometer Camera observing at 870 $\mu$m (345 GHz). The APEX telescope has a full width at half-maximum (FWHM) beam size of  19$^{\prime\prime}$ at this  wavelength.

To trace the ionized gas of M16, we also used the Digitized Sky
Survey \citep[DSS1,][]{Reid1991}  and 90 cm radio continuum emission archival data. The 90 cm data is obtained from multiconfiguration Very Large Array survey of the Galactic Plane with a resolution of 42$^{\prime\prime}$ \citep{Brogan2006}. 

\begin{figure*}
\centering
\includegraphics[width = 0.65 \textwidth]{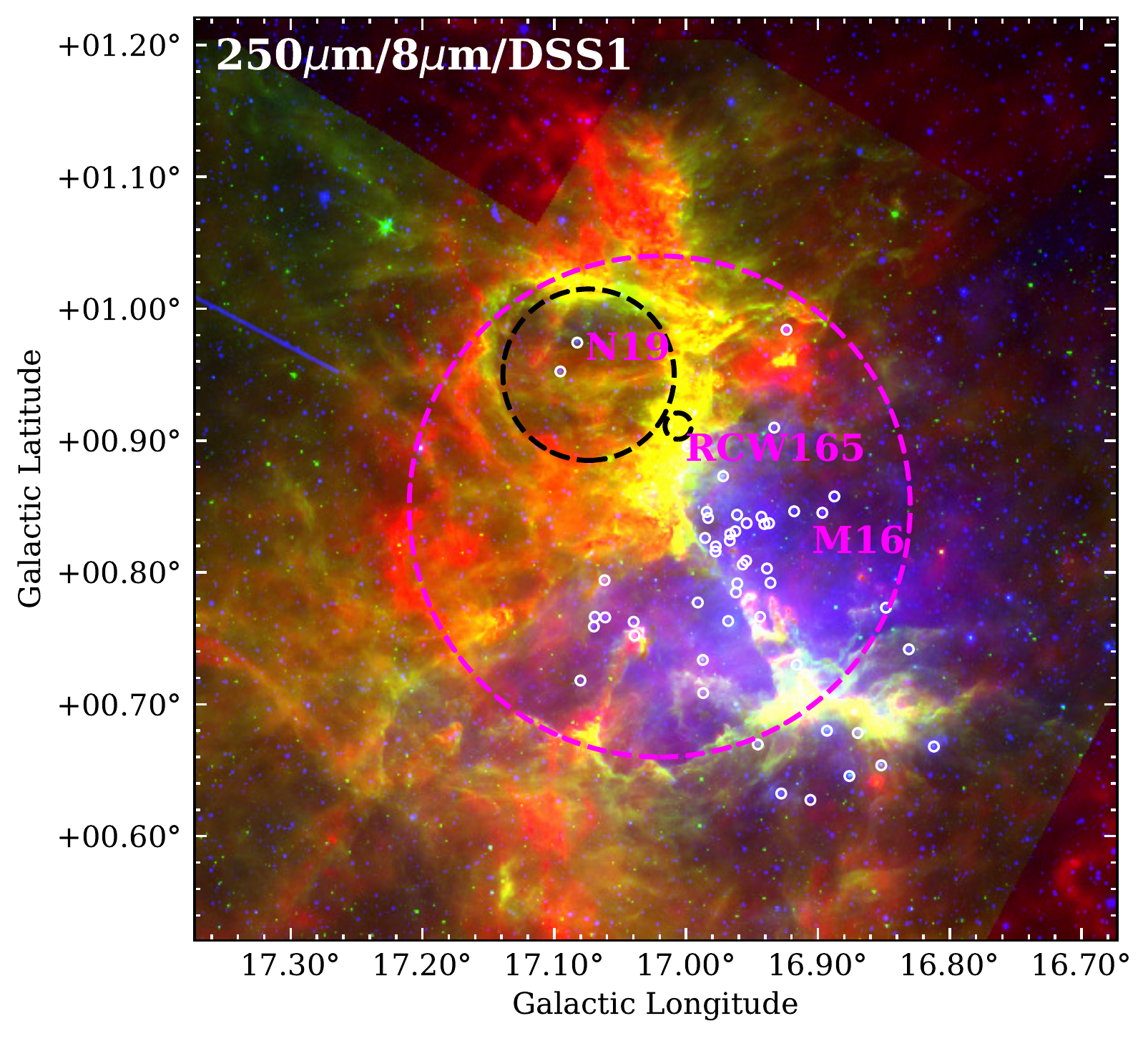}
\vspace{-4mm}
\caption{Three-color composite image of M16 using {\it Herschel} 250 $\mu$m (red), {\it Spitzer}-IRAC 8 $\mu$m (green), and DSS1-red (blue). The white circles indicate the positions of O and B stars  \citep{Evans2005}. The pink and black dashed circles mark the 250 $\mu$m ring-like structure (M16 \HII region), and bubbles N19 and RCW165, respectively.}
\label{Fig:DSS-8UM}
\end{figure*}

\section{Results}
\label{sect:results}
\subsection{Optical, infrared, and radio continuum images}
\label{sect:Optical}
Figure \ref{Fig:DSS-8UM} shows a composite three-color image for  the M16 \HII region. The three bands shown are the {\it Herschel} 250 $\mu$m  (in red), {\it Spitzer} 8 $\mu$m (in green), and optical DSS1-red (in blue).  The 8 $\mu$m emission  displays several pillars with the heads faced toward the central part of M16. In addition, the 8 $\mu$m emission also shows a small ring-like structure located on the northeastern edge of M16. Generally, the 8 $\mu$m band can be used to trace  polycyclic aromatic hydrocarbon (PAH) emission and delineates the \HII region boundaries  \citep{Pomares2009}. Compared with the bubble catalog of \citet{Churchwell2006}, we find that the ring-like structure is related to bubble N19. While the 250 $\mu$m emission originates from cool dust \citep{Anderson2012}. The cool dust emission  (red color) also shows some small filament-like structures, which are mainly located on the northeastern edge of M16. Moreover, the 250 $\mu$m emission shows a ring-like structure, which is delineated using  a pink dashed circle. The optical DSS1-red image presents the ionized-gas emission of M16, which is enclosed by the PAH and cool dust emission with an opening towards the northwest. 

\begin{figure}
\centering
\includegraphics[width = 0.46 \textwidth]{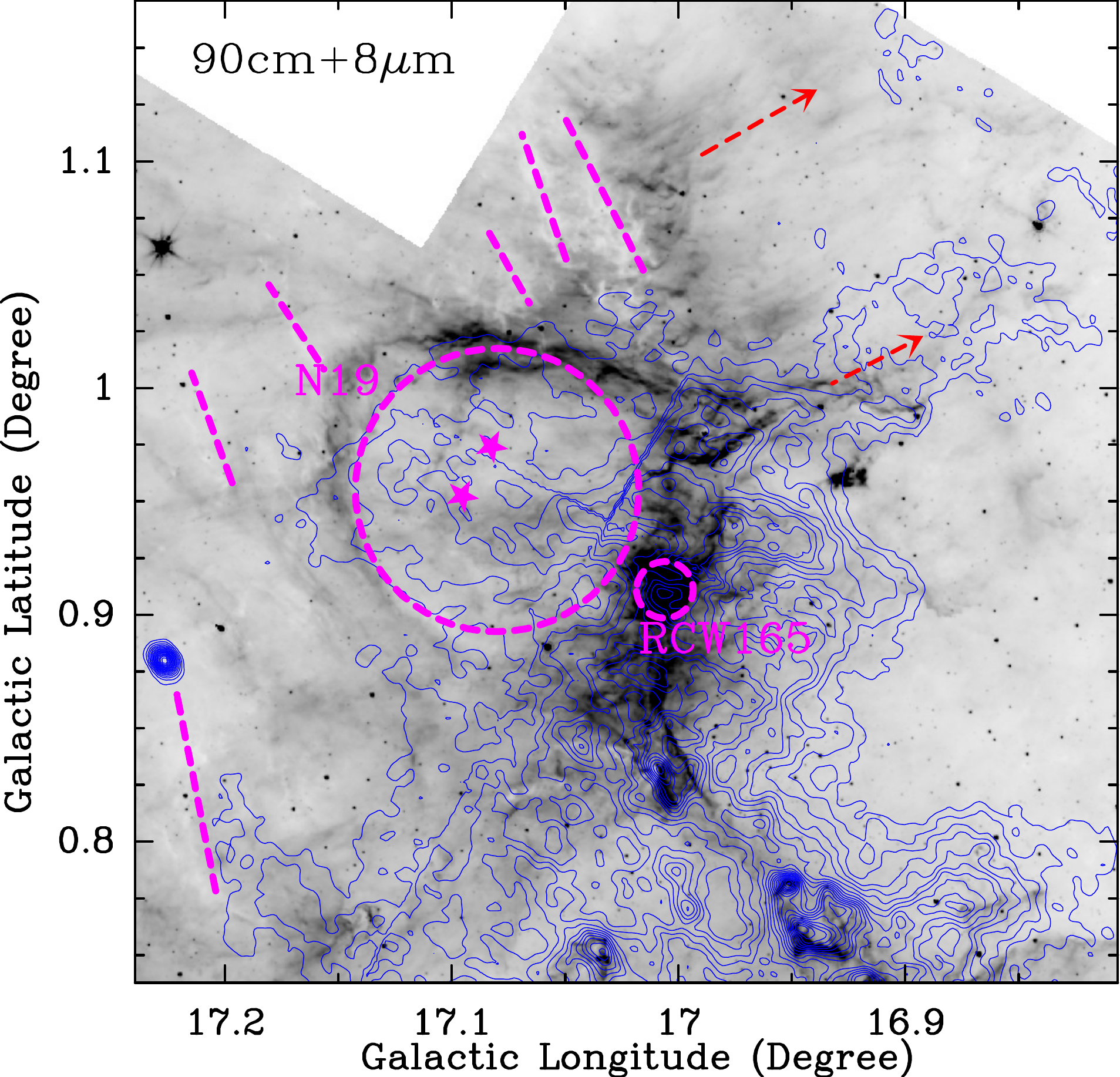}
\vspace{-2mm}
\caption{The 90 cm emission (blue contours) superimposed on the {\it Spitzer}-IRAC 8 $\mu$m emission (grayscale). The pink dashed circles represent bubbles N19 and RCW165. The contour levels start from 0.021 Jy/beam (3$\rm\sigma$) in steps of 3$\rm \sigma$. The pink dashed lines indicate the dark IR clouds \citep{Peretto2009}, while the red arrows show two gas flows. The pink stars mark the positions of the two ionizing stars of N19  \citep{Evans2005}.}
\label{Fig:8um-90cm}
\end{figure}

\begin{figure}
\centering
\includegraphics[width = 0.46 \textwidth]{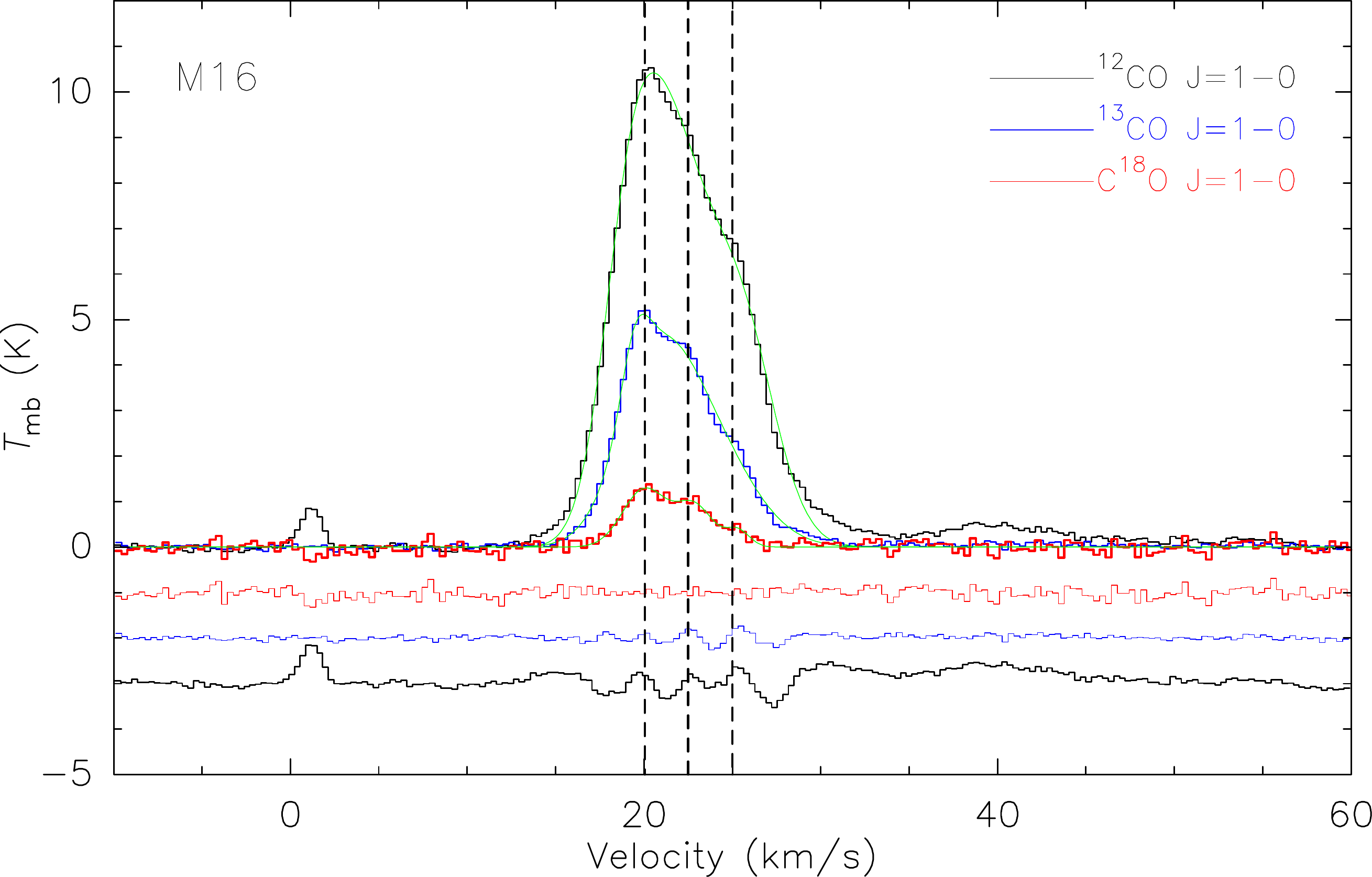}
\caption{Averaged spectra of $^{12}$CO $J$=1-0, $^{13}$CO $J$=1-0, and C$^{18}$O $J$=1-0  over the entire M16. We note that the intensities of $^{13}$CO $J$=1-0, and C$^{18}$O $J$=1-0 are multiplied by a factor of 2 and 5 for clarity, respectively. The green lines represent the Gaussian fitted
lines,  with the resulting residuals shown below. The black vertical dashed lines mark the different velocity components.}
\label{Fig:spectrum}
\end{figure}

\begin{figure*}
\centering
\includegraphics[width = 0.96 \textwidth]{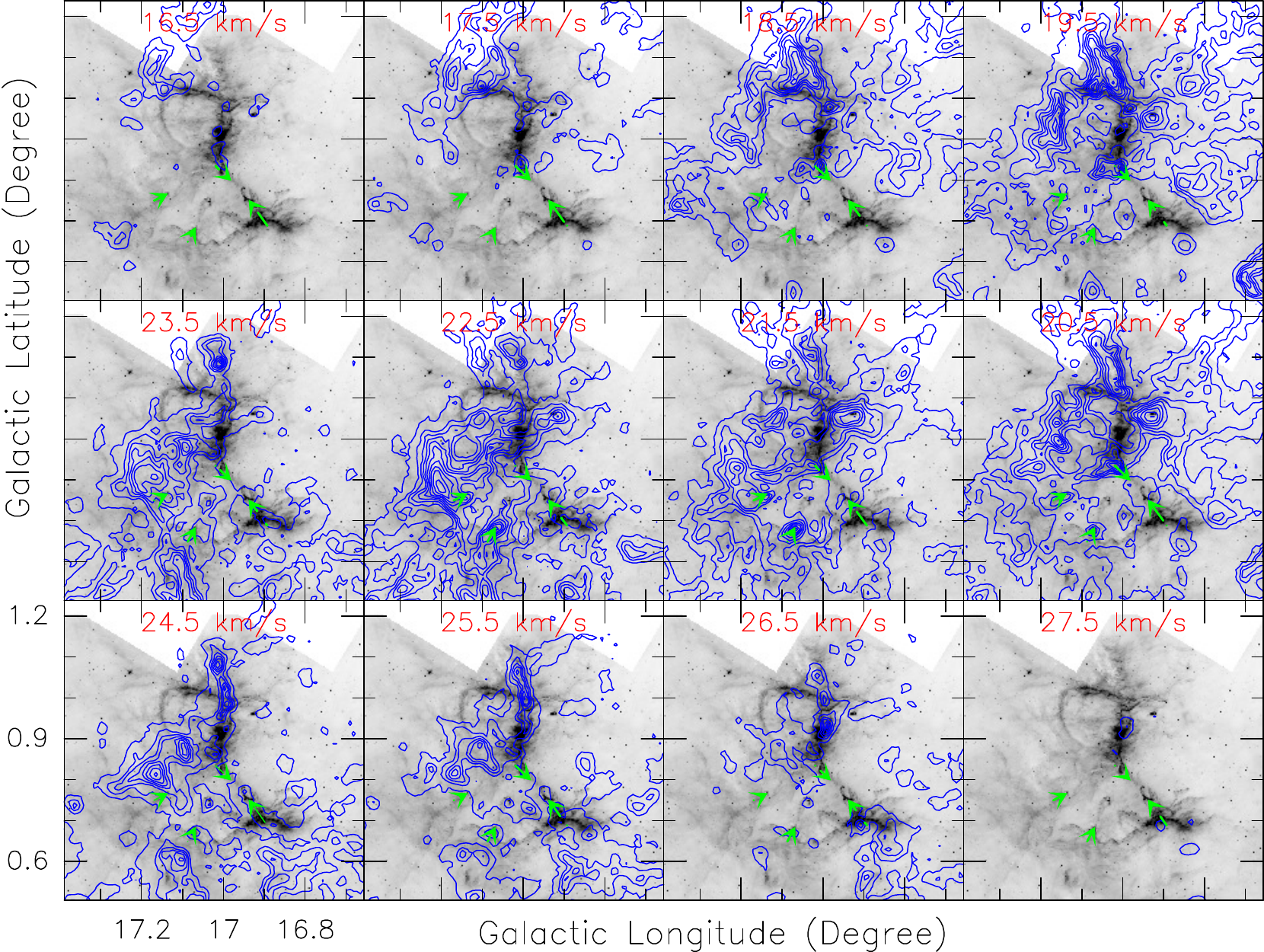}
\caption{$^{13}$CO $J$=1-0 channel maps from the integrated emission in step of 1 km s$^{-1}$ overlaid on the Spitzer-IRAC 8 $\mu$m emission (gray). Central velocities are indicated in each image. The green arrows represent the pillars.}
\label{Fig:8um-cmap}
\end{figure*}

\begin{figure*}
\centering
\includegraphics[width = 0.40 \textwidth]{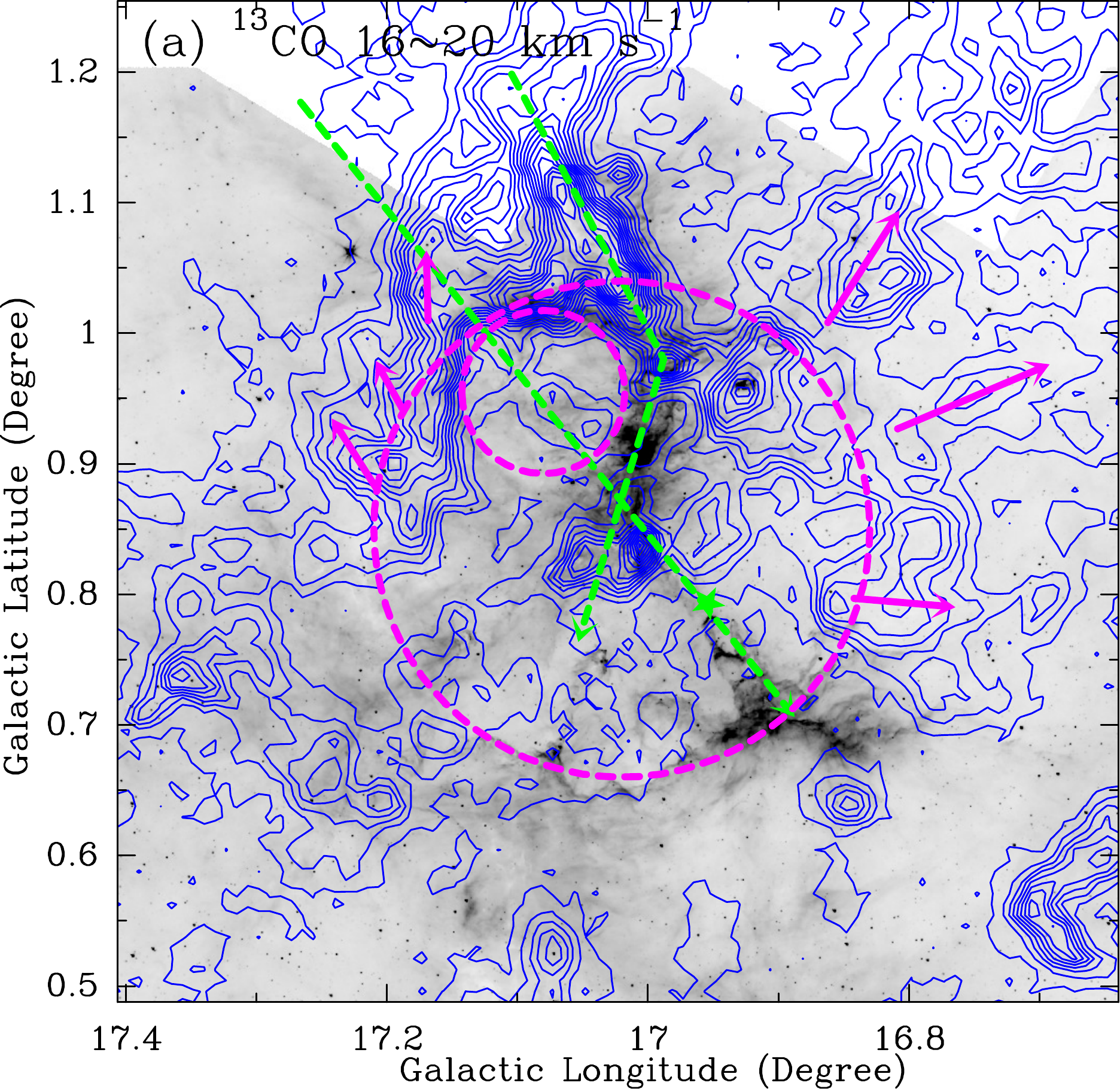}
\includegraphics[width = 0.40 \textwidth]{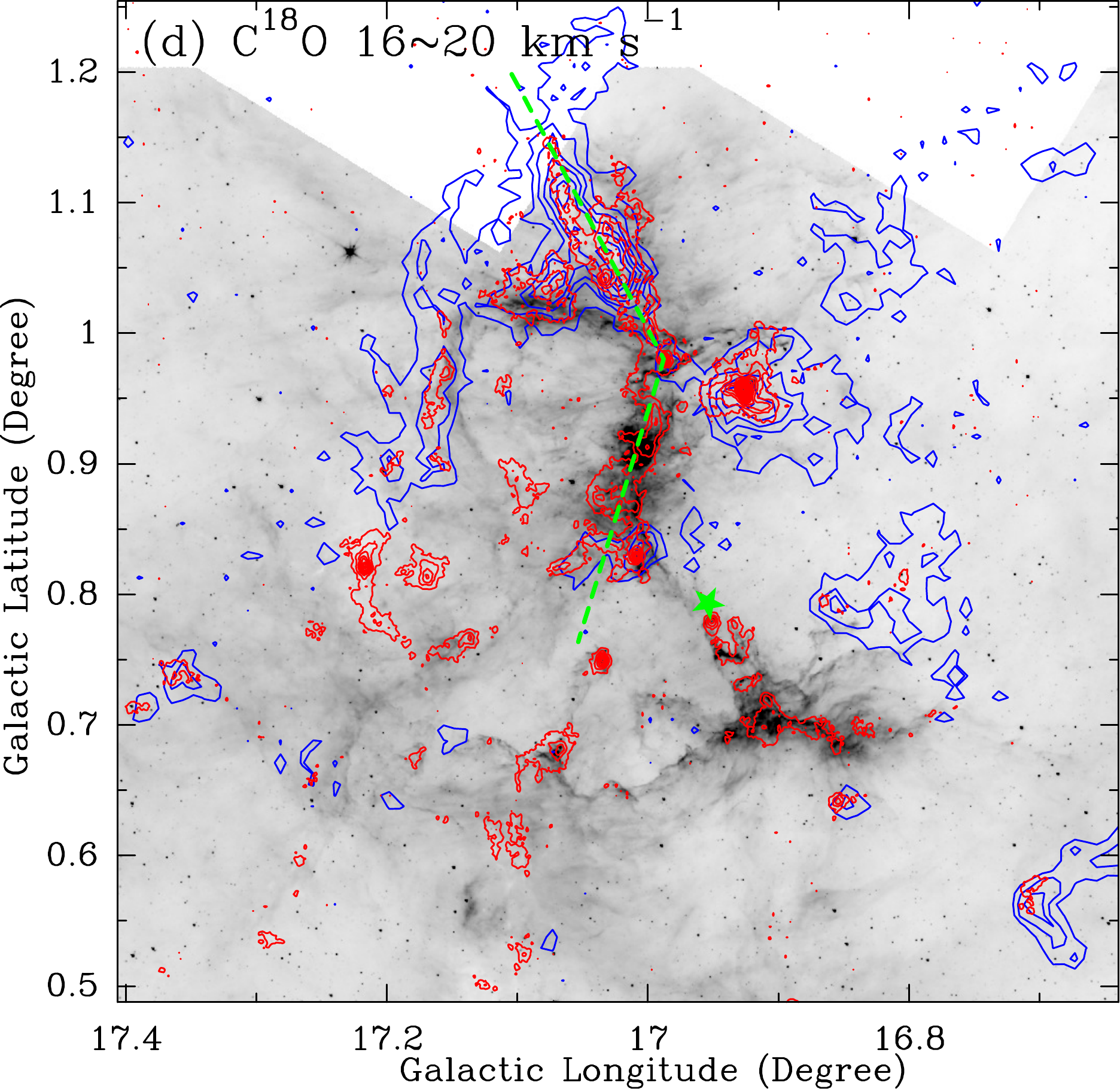}
\includegraphics[width = 0.40 \textwidth]{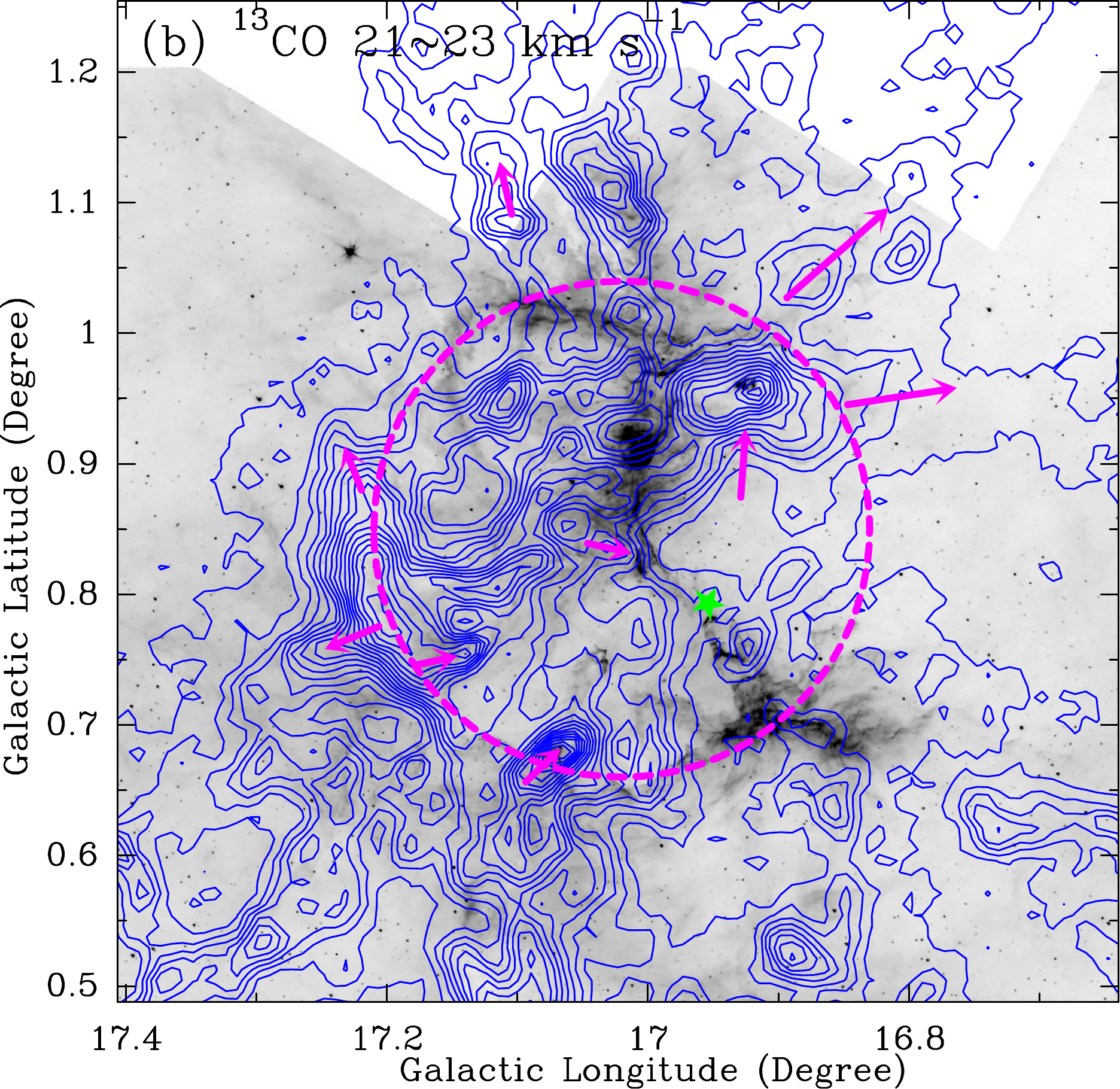}
\includegraphics[width = 0.40 \textwidth]{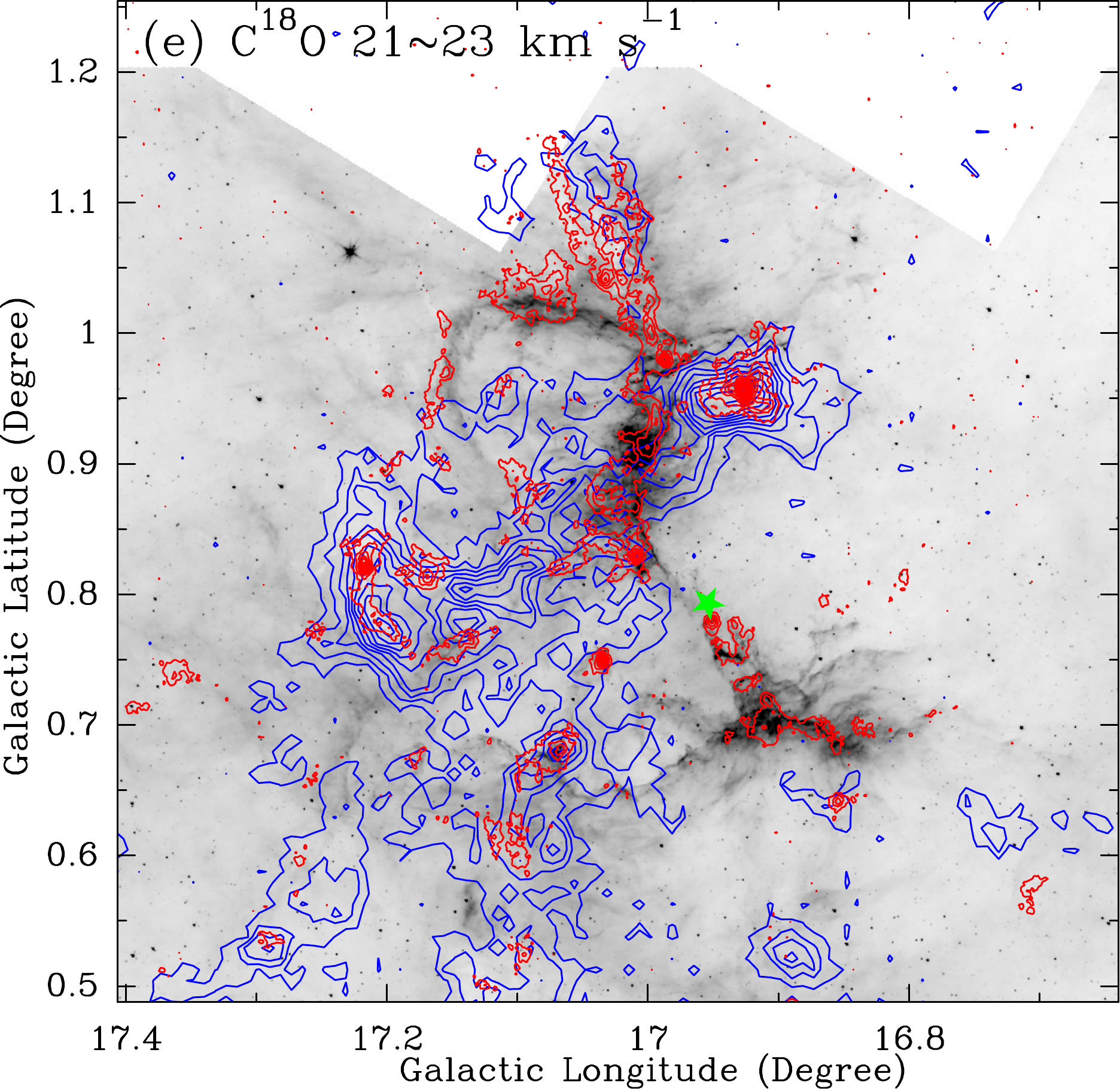}
\includegraphics[width = 0.40 \textwidth]{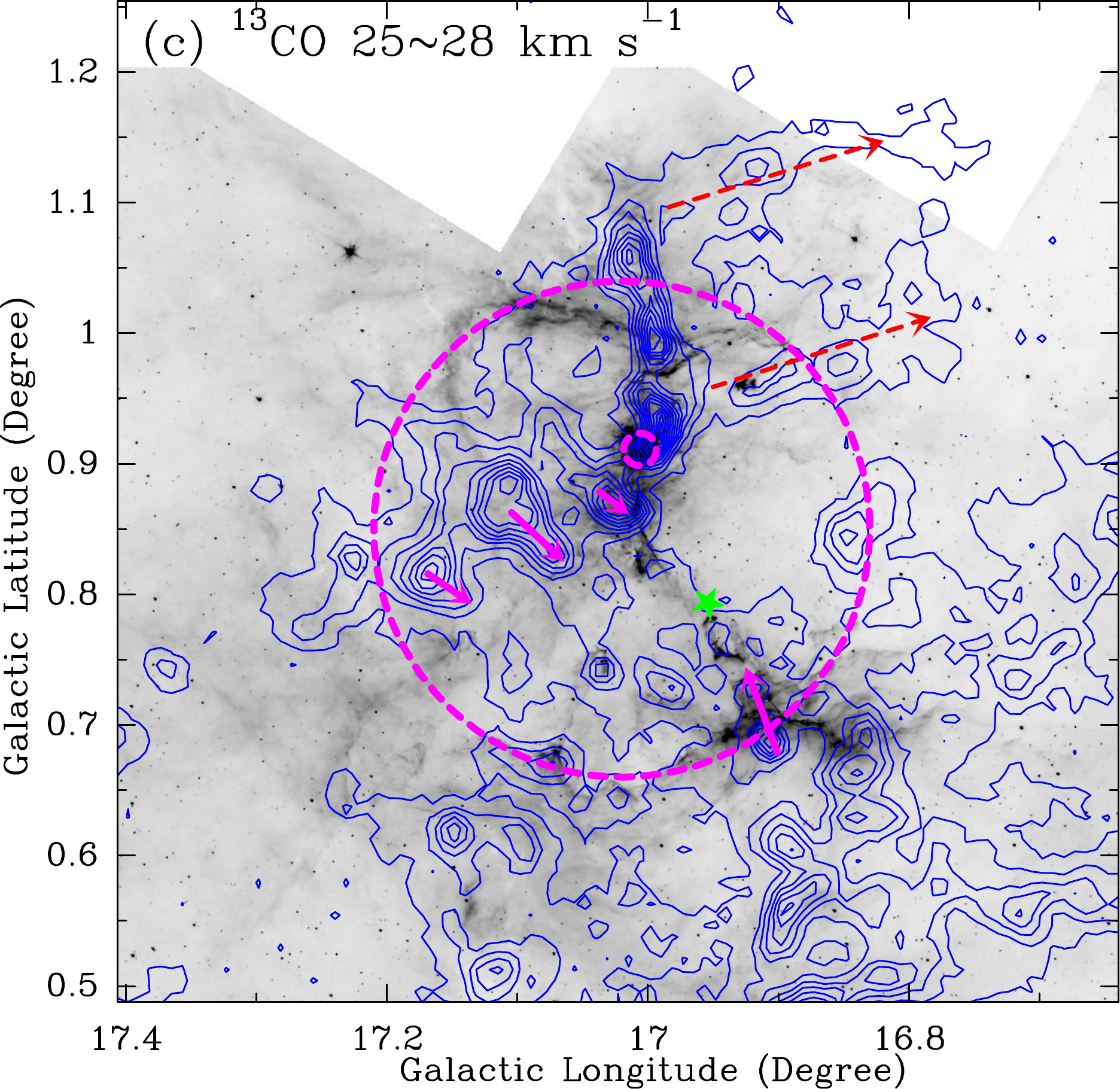}
\includegraphics[width = 0.40 \textwidth]{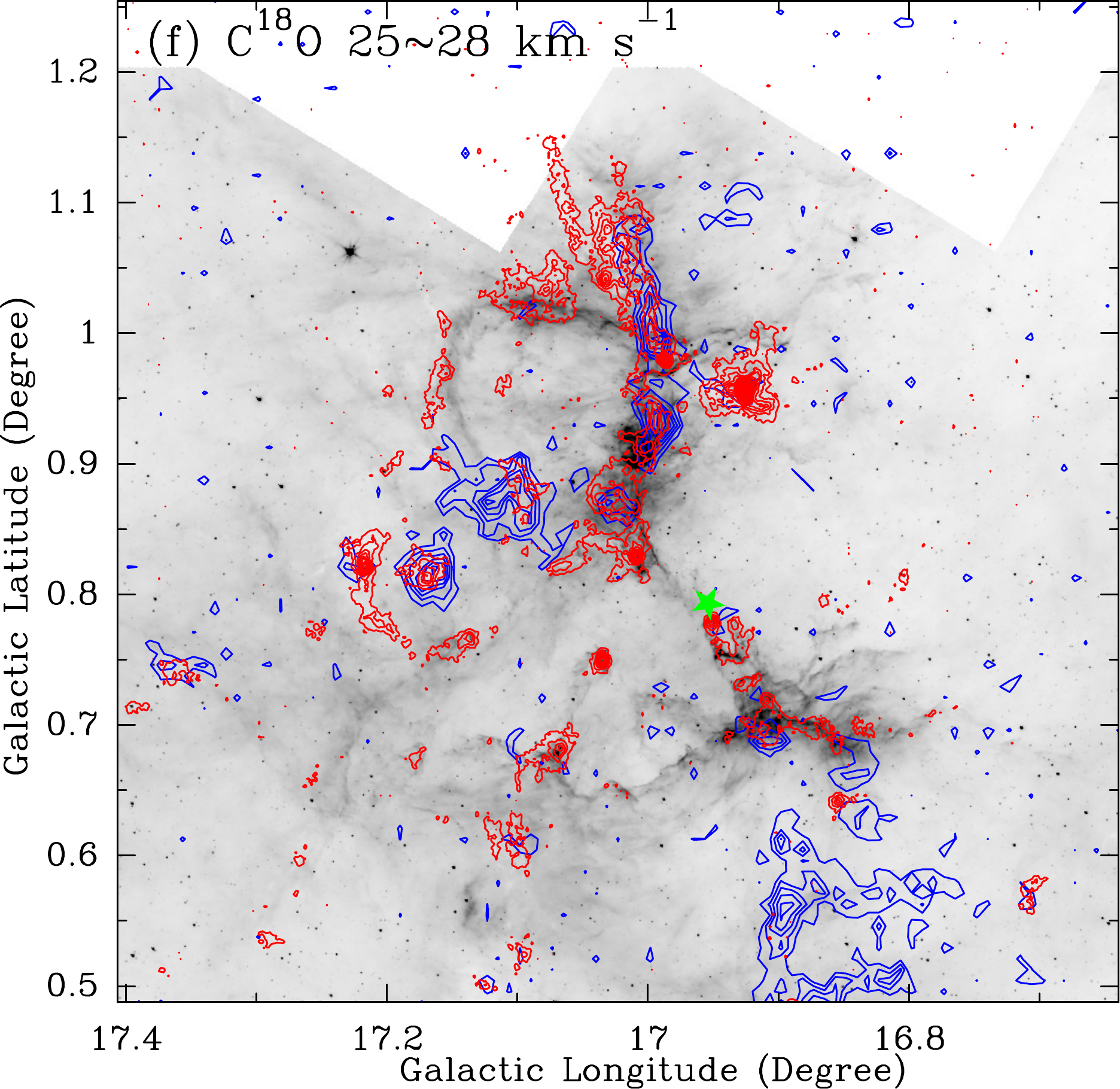}
\caption{(a)(b)(c) panels: $^{13}$CO integrated intensity maps (blue contours) superimposed on the {\it Spitzer}-IRAC 8 $\mu$m emission (gray). The integrated-velocity ranges are shown on top of each image. The pink arrows mark the pillars, while the red  dashed arrows indicate two gas flows. The blue contour levels start from 0.2 K km s$^{-1}$(5$\rm\sigma$) in a step of 1.6 K km s$^{-1}$.  The pink dashed circles represent M16, N19, and RCW165. The green star marks the NGC6611 cluster ionizing M16.   (d)(e)(f) panels: C$^{18}$O integrated intensity images overlaid with the ATLASGAL 870 $\mu$m (red contours) and 8 $\mu$m emission (gray).  The blue contour levels start from 0.15 K km s$^{-1}$(5$\rm\sigma$) in a step of 0.75 K km s$^{-1}$, while the red contour levels start from 0.06 Jy/beam (3$\rm\sigma$) in a step of 0.3 Jy/beam.  The green dashed lines mark the directions of position--velocity diagrams in Figure \ref{Fig:PV}. }
\label{Fig:CO-8UM}
\end{figure*}

Figure \ref{Fig:8um-90cm} shows the 90 cm emission map (blue contours) overlaid with the 8 $\mu$m emission (grayscale).  The 90 cm continuum emission is used to trace  the ionized gas. In Figure \ref{Fig:8um-90cm}, the ionized gas is mainly coincident with the M16 \HII region. \citet{Evans2005} identified  the O and B stars in the whole M16 region, as shown in Figure \ref{Fig:DSS-8UM}. Two of these stars, which may be the ionizing stars of N19, are observed  towards the bubble.  Moreover, the ionized gas of N19 clearly shows a gas flow towards the west, which is spatially associated with the PAH  emission.  The RCW165 \HII region  \citep{Dubout-Crillon1976} is located on a north-south (NS) filament. Adjacent to the northeast of M16, we find some smaller dark filaments, marked in the pink dashed lines in Figure \ref{Fig:8um-90cm}, which are associated with smaller IRDCs \citep{Peretto2009}. 

\begin{figure*}
\centering
\includegraphics[width = 0.36 \textwidth]{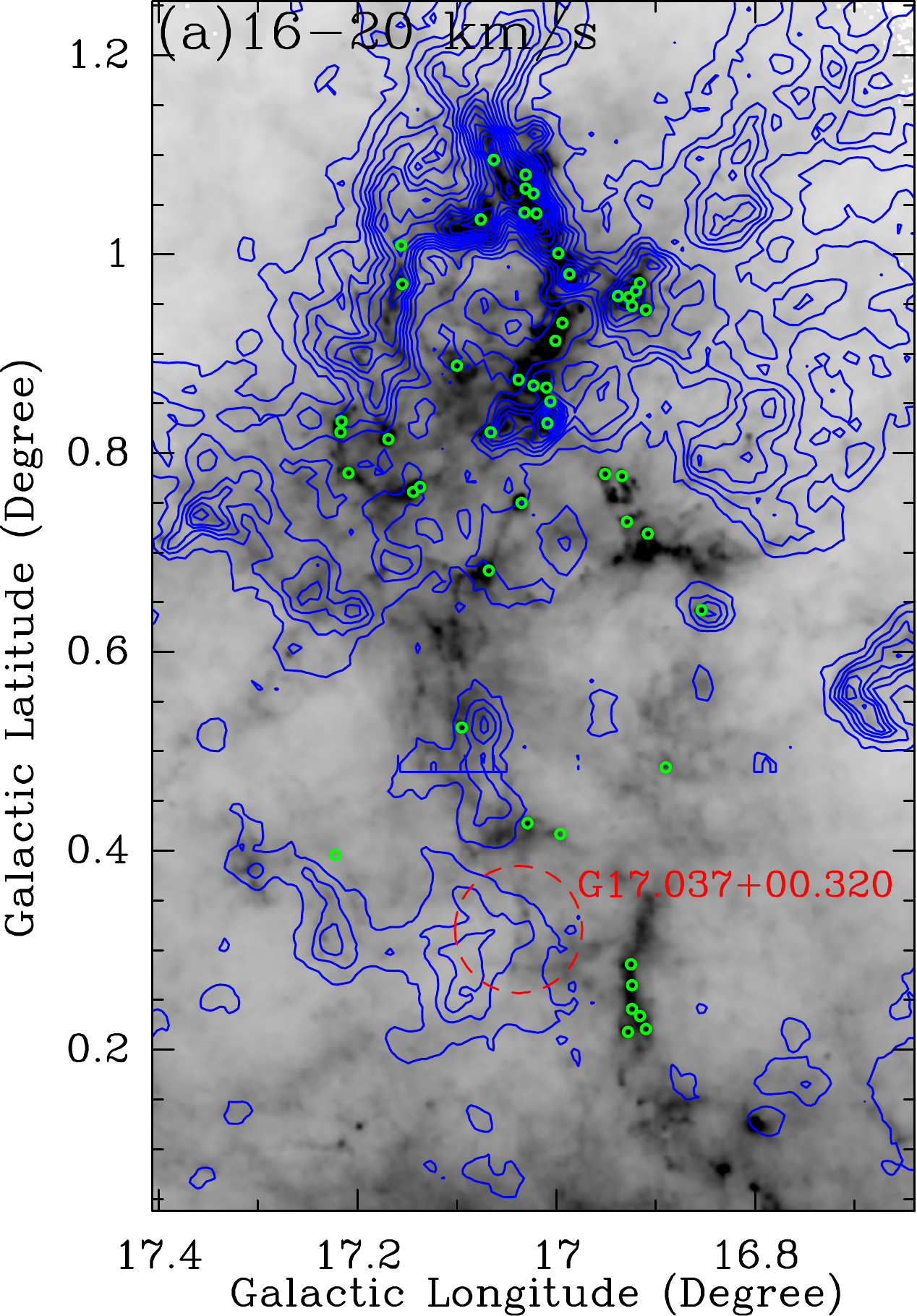}
\includegraphics[width = 0.36 \textwidth]{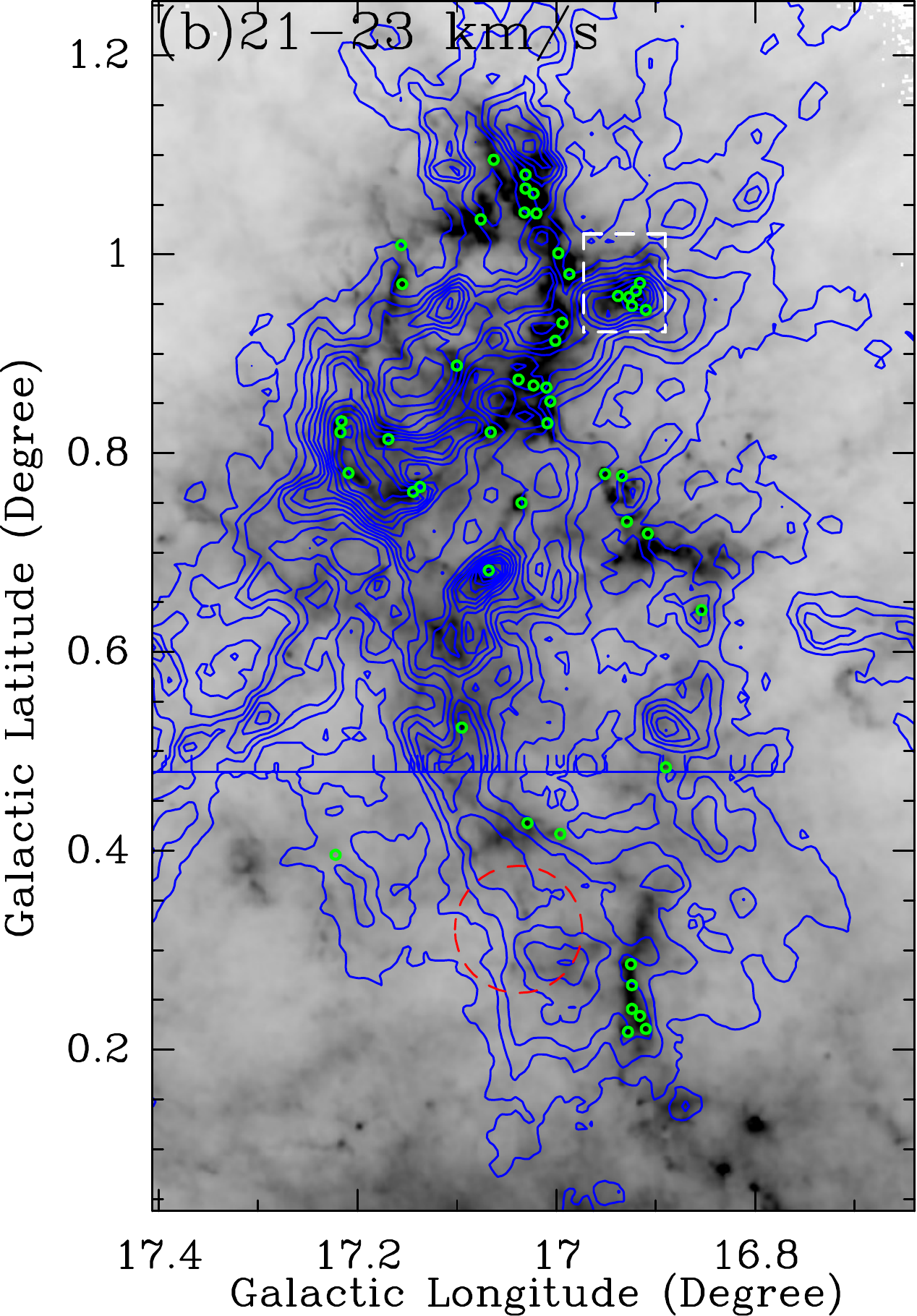}
\includegraphics[width = 0.36 \textwidth]{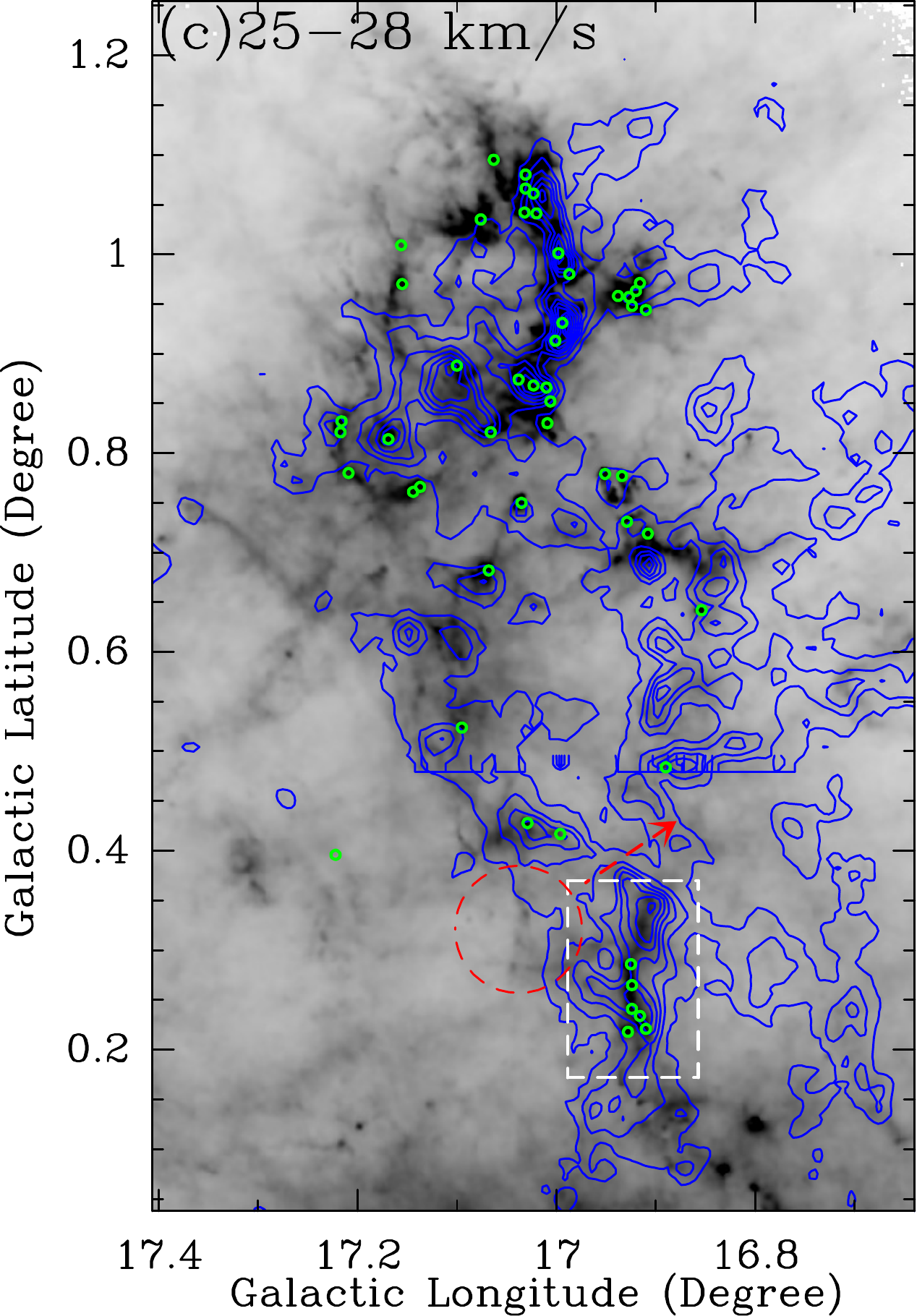}
\includegraphics[width = 0.36 \textwidth]{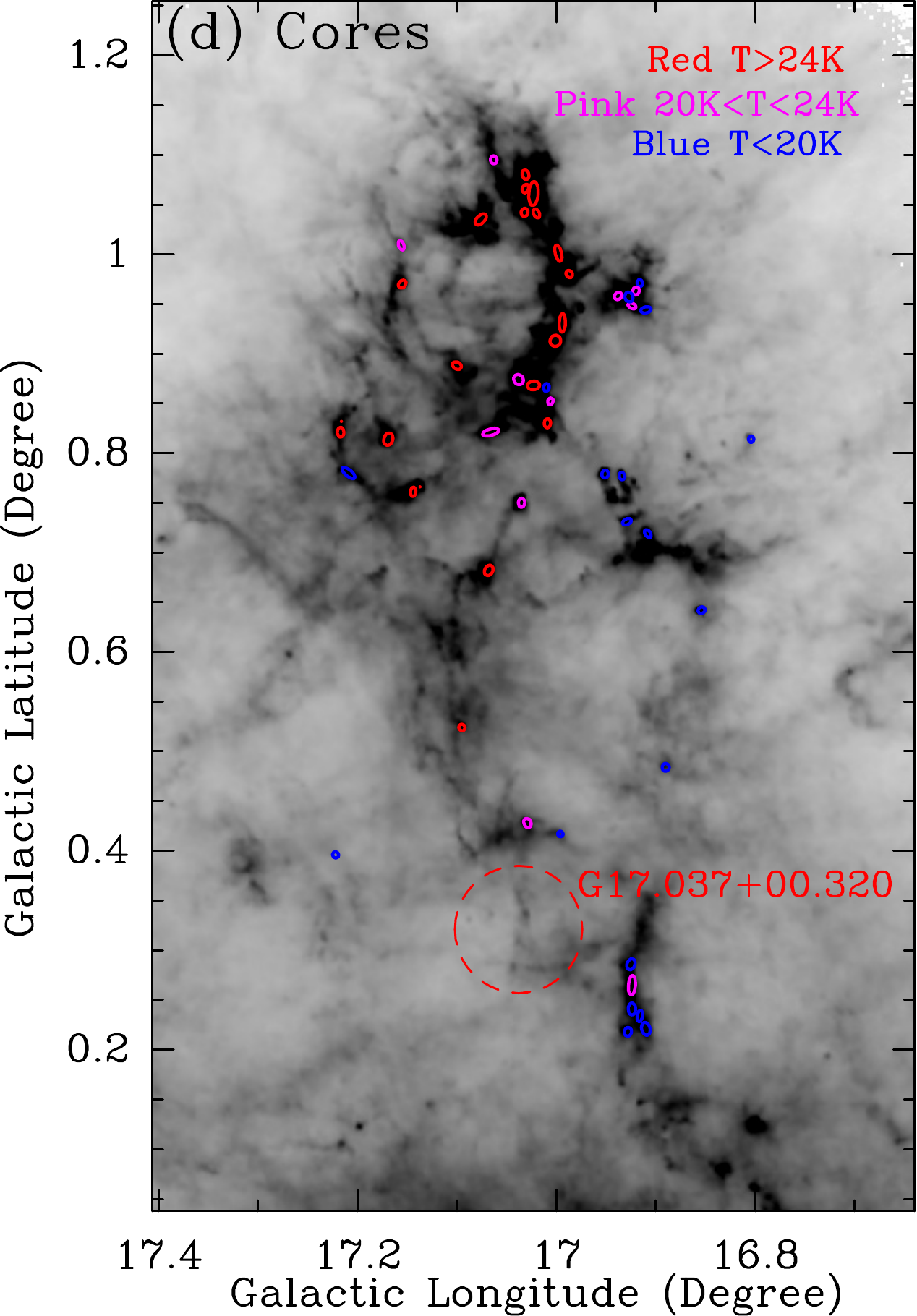}
\caption{(a)(b)(c): The  $^{13}$CO $J$=1-0 emission in blue contours overlaid on the $Herschel$ 250 $\mu$m emission in grayscale. The integrated-velocity ranges are shown on top of each image. The green circles represent the 870 $\mu$m dust cores, while the red  dashed circle shows the position and size of G17.037+0.320 \HII region  from \citet{Anderson2014}. The red arrow may presents the shock direction.  The two white dashed  boxes indicate the compact region with six dust cores. (d): The ellipses also mark the positions, sizes, and  position angles of the 870 $\mu$m dust cores, which are given in Table 1.  The different colors mark the cores with different temperatures ($T$). If $T$ $>$ 24 K, the cores are marked in red, for 10 K $<$ $T$ $<$ 24 K and $T$ $<$ 10 K in pink and blue, respectively.}
\label{Fig:W37}
\end{figure*}

\begin{figure*}
\centering
\includegraphics[width = 0.96 \textwidth]{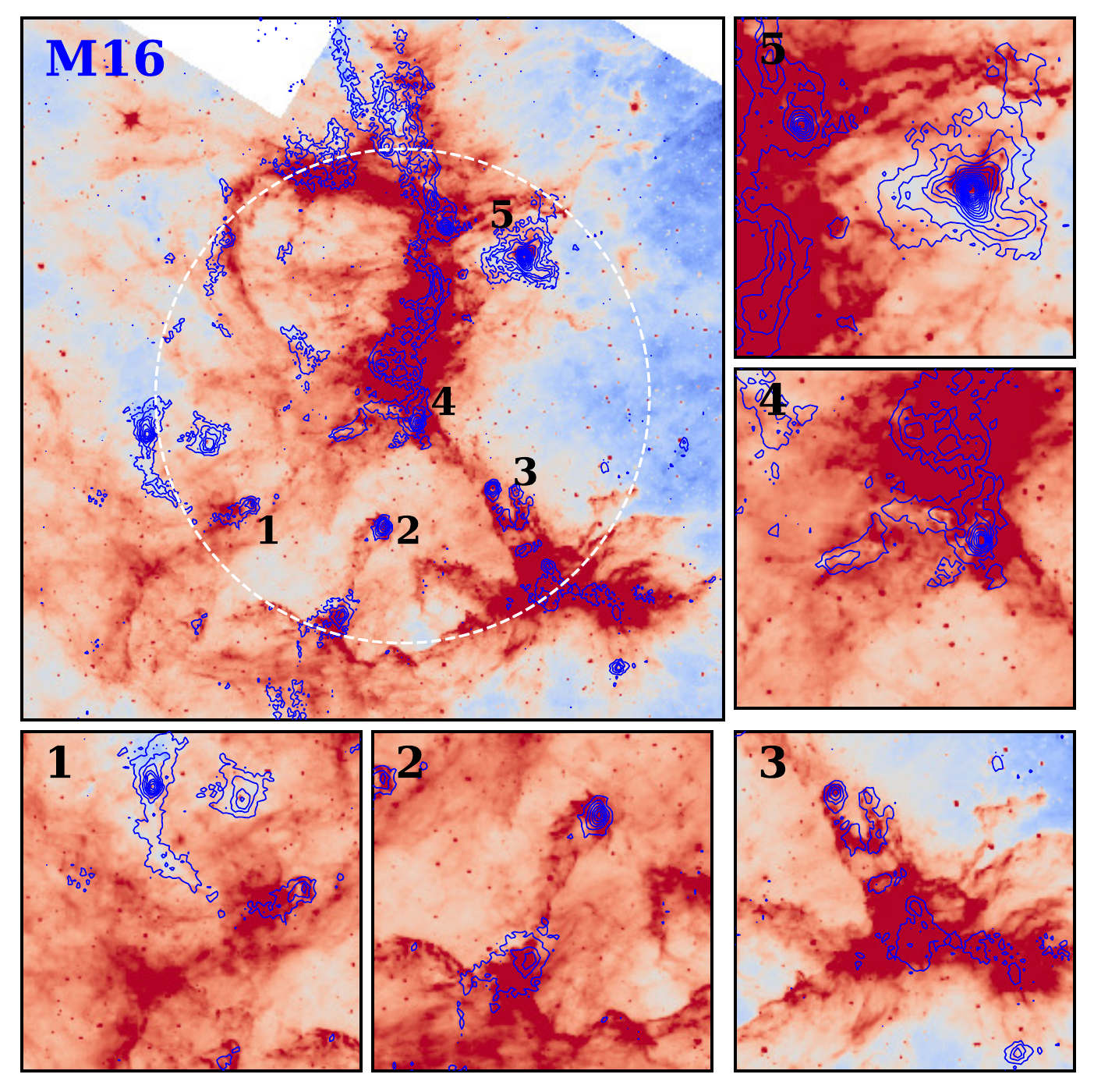}
\caption{The 870 $\mu$m emission (blue contours) is superimposed on the {\it Spitzer}-IRAC 8
$\mu$m emission (red). The typical pillars and compact cores also are shown in the zoomed image.  The white dashed circle outlines the  M16 \HII region. The blue contour levels start from 0.06 Jy/beam (3$\rm\sigma$) in steps of 0.3 Jy/beam}
\label{Fig:Pillar}
\end{figure*}

\subsection{Carbon monoxide molecular emission} 
The dust continuum emission only gives the distribution of the projected 2D gas around the M16 \HII region. To investigate in detail the gas structure  associated with the M16 \HII region, we use $^{12}$CO $J$=1-0, $^{13}$CO $J$=1-0, and C$^{18}$O $J$=1-0 lines to trace the molecular gas.  Figure \ref{Fig:spectrum} shows the averaged spectra of $^{12}$CO $J$=1-0, $^{13}$CO $J$=1-0, and C$^{18}$O $J$=1-0 over  the entire M16 \HII region. All the $^{12}$CO $J$=1-0, $^{13}$CO $J$=1-0, and C$^{18}$O $J$=1-0 spectra show three peaks whose position is at 20.0$\pm$0.4, 22.5$\pm$0.4, and 25.0$\pm$0.4 km s$^{-1}$,  which probably indicate  different molecular components along the line of sight. Because the three components in the $^{12}$CO $J$=1-0 and $^{13}$CO $J$=1-0 spectra are too close, we are not able to see their distinct peaks, except for a few protrusions. The most obvious result is that the C$^{18}$O $J$=1-0 spectrum shows three clear peaks. Using the $^{12}$CO $J$=1-0 data with an angular resolution of 180$^{\prime\prime}$, \citet{Nishimura2017} also found three velocity components peaked at 19.5, 23.0, and 25.0 km s$^{-1}$  through a latitude-velocity diagram for the M16 region. Within the error range, our measured peak velocities for the three components are equal to those obtained by \citet{Nishimura2017}.
Compared with the $^{13}$CO $J$=1-0 and C$^{18}$O $J$=1-0 spectral profiles,  the optically thick $^{12}$CO line shows that there are two other velocity components, whose peak velocities are at 1.5 and 40.0 km s$^{-1}$, respectively. These two velocity components may be related to background Galactic weak gas emission. Hence, the $^{13}$CO line is more suited to trace relatively dense gas. Using channel maps of the $^{13}$CO line,   we further check the gas components which are associated with the M16 \HII region.

Figure \ref{Fig:8um-cmap} shows the $^{13}$CO $J$=1-0 channel maps overlaid on the 8 $\mu$m emission, whose velocity ranges from 16 to 28 km s$^{-1}$ in steps of 1 km s$^{-1}$. Based on the morphology of the molecular gas associated with the previous identified bubble/\HII regions, pillars, and  filaments, we find three velocity components. Component 1, which is located in velocity ranges  from 16 to 24 km s$^{-1}$, is mainly  consistent with N19 and a massive gas clump identified in our data. The clump is situated in the northwest region of M16.  Component 2,  observed from 20 to 24 km s$^{-1}$, is associated with some pillars.  Component 3, which is located in the velocity ranges from 23 to 28 km s$^{-1}$, is mainly  consistent with the NS filament. We suggest that these three components overlap and are connected to each other in velocity and space.  To show the gas structure associated with each velocity component, we use the smaller integrated-velocity ranges to make the integrated-intensity images of $^{13}$CO $J$=1-0. The three emissions are shown in Fig. \ref{Fig:CO-8UM}(a)(b)(c), overlaid on the 8 $\mu$m emission. The integrated-velocity ranges are shown on top of each map. The $^{13}$CO $J$=1-0 emission of component 1 exhibits a small ring-like shape with an opening at the southwest, which just surrounds bubble N19. In addition, from component 1 to component 3, we find that some $^{13}$CO $J$=1-0 emission with pillars is distributed around M16, and the other emission towards the interior of the M16 \HII region. In component 3, we find two gas flows marked with red arrows, which are also seen in Figure \ref{Fig:8um-90cm}. Figure \ref{Fig:CO-8UM}(d)(e)(f) shows the C$^{18}$O integrated-intensity images of the different velocity components overlaid with the 870 $\mu$m (red contours) and 8 $\mu$m emission (gray).  The spatial distribution of the C$^{18}$O $J$=1-0 emission is similar to that of the 870 $\mu$m dust emission. The 870 $\mu$m emission also displays the NS filament.

Additionally, to construct a large-scale CO picture for  the M16 \HII region, we also use the GRS $^{13}$CO $J$=1-0 data.  Figure \ref{Fig:W37} shows the $^{13}$CO $J$=1-0 integrated-intensity maps overlaid on the  {\it Herschel} 250 $\mu$m emission. Each velocity component coincides well with its counterpart in the 250 $\mu$m emission. Both emissions show a filamentary structure (large-scale filament). The  G17.037+0.320 HII region \citep{Anderson2014} is shown in Figure \ref{Fig:W37}. This  expansion of the \HII region might have divided the large-scale filament into  two parts.

\subsection{The selected cores}
Using the ATLASGAL catalog \citep{Contreras2013}, we extract 51 dust cores in the large-scale  filament. These cores are mainly distributed along the large-scale filament in Figure \ref{Fig:W37}, but more cores are located on the northern part of the large-scale filament. Since the 870 $\mu$m emission traces the distribution of dense cold dust, the northern part is likely to be the densest in the large-scale filament, which is associated with the M16 \HII region.  We also present a zoomed image for the northern part of the large-scale filament to show the relation between M16 and these cores, as shown in Figure \ref{Fig:Pillar}.  There is a large compact clump located at the northwest of M16.  We extracted six dust cores in the large clump from the catalog of \citet{Contreras2013}. Particularly, there are some cores that are situated in the heads of the pillars. Figure \ref{Fig:Pillar} also shows four typical pillars associated with the 870 $\mu$m cores. The parameters of the selected cores  are listed in Table 1.

\begin{figure*}
\centering
\includegraphics[width = 0.46 \textwidth]{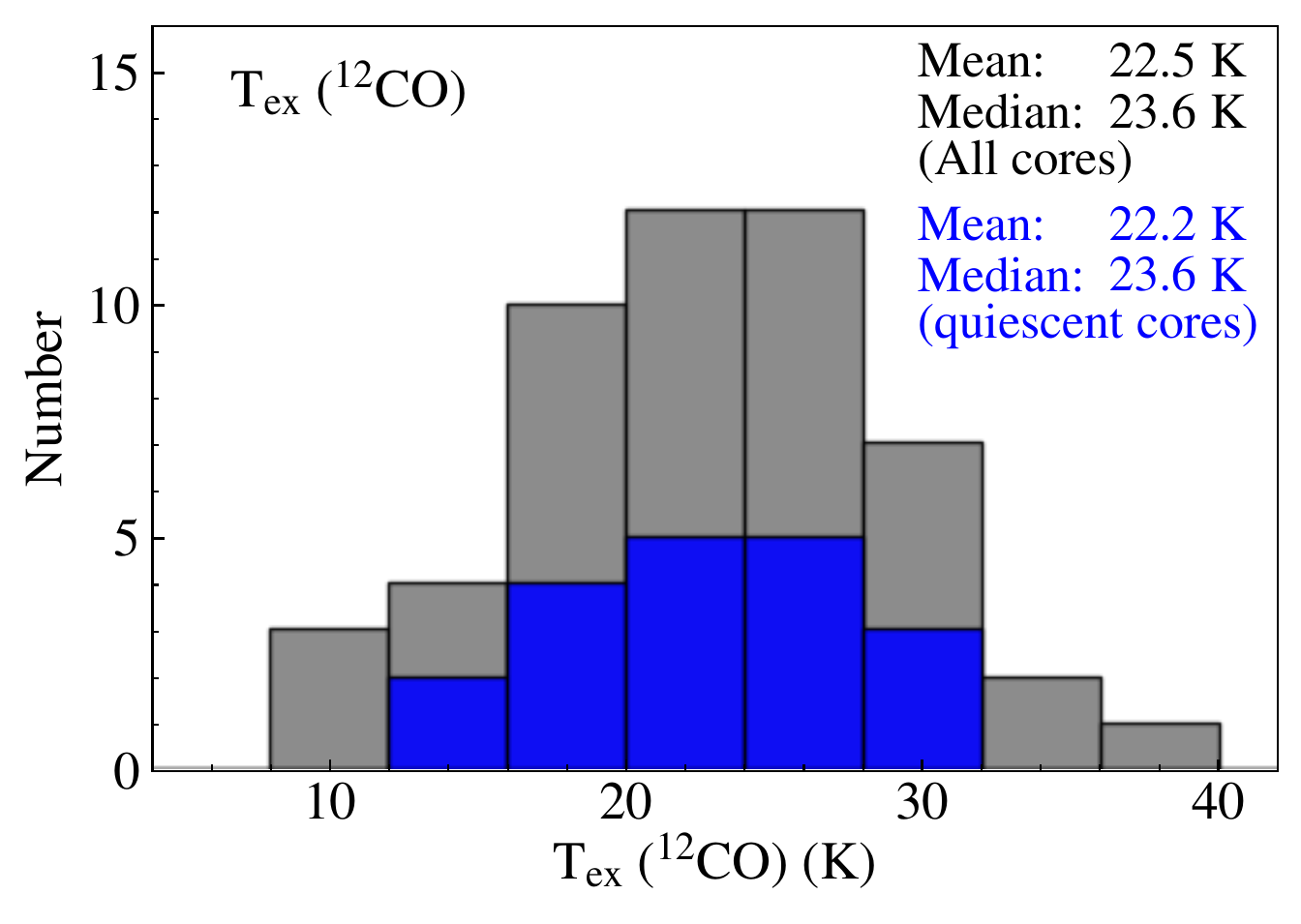}
\includegraphics[width = 0.47\textwidth]{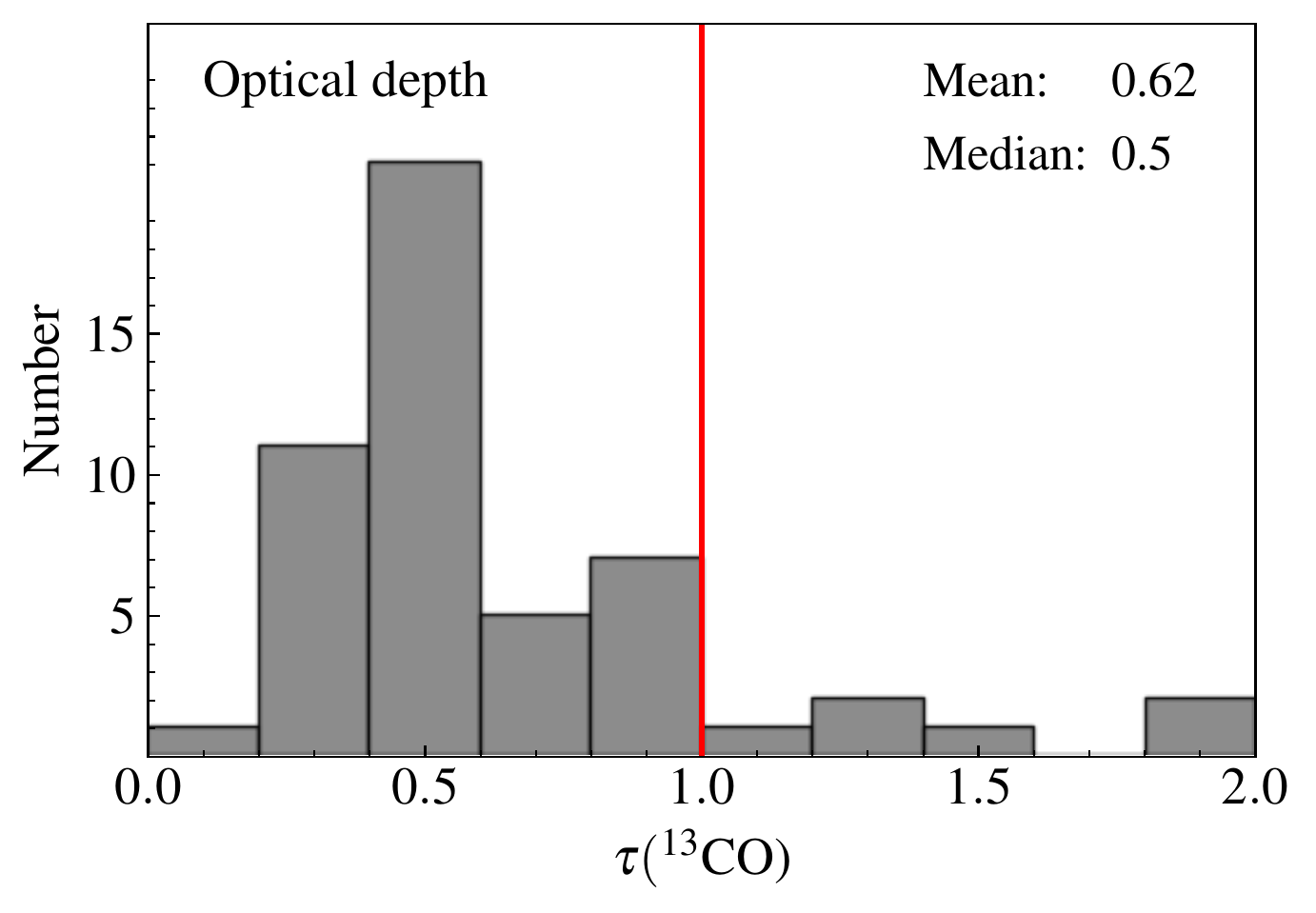}
\includegraphics[width = 0.46 \textwidth]{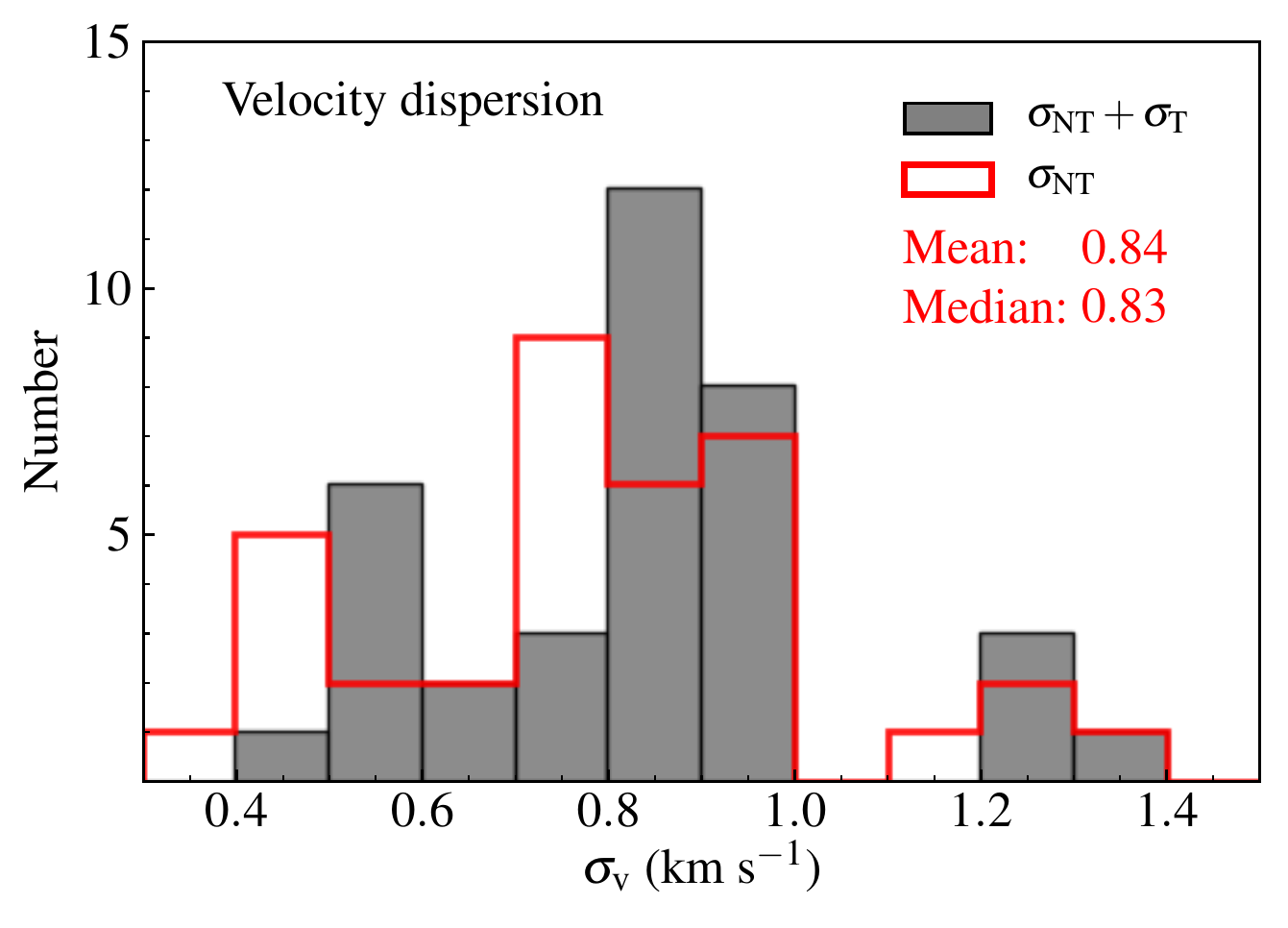}
\vspace{-4mm}
\caption{Distributions of excitation temperature, optical depth, and velocity dispersion for the selected cores.}
\label{Fig:Tex}
\end{figure*}

\begin{figure*}
\centering
\includegraphics[width = 0.45 \textwidth]{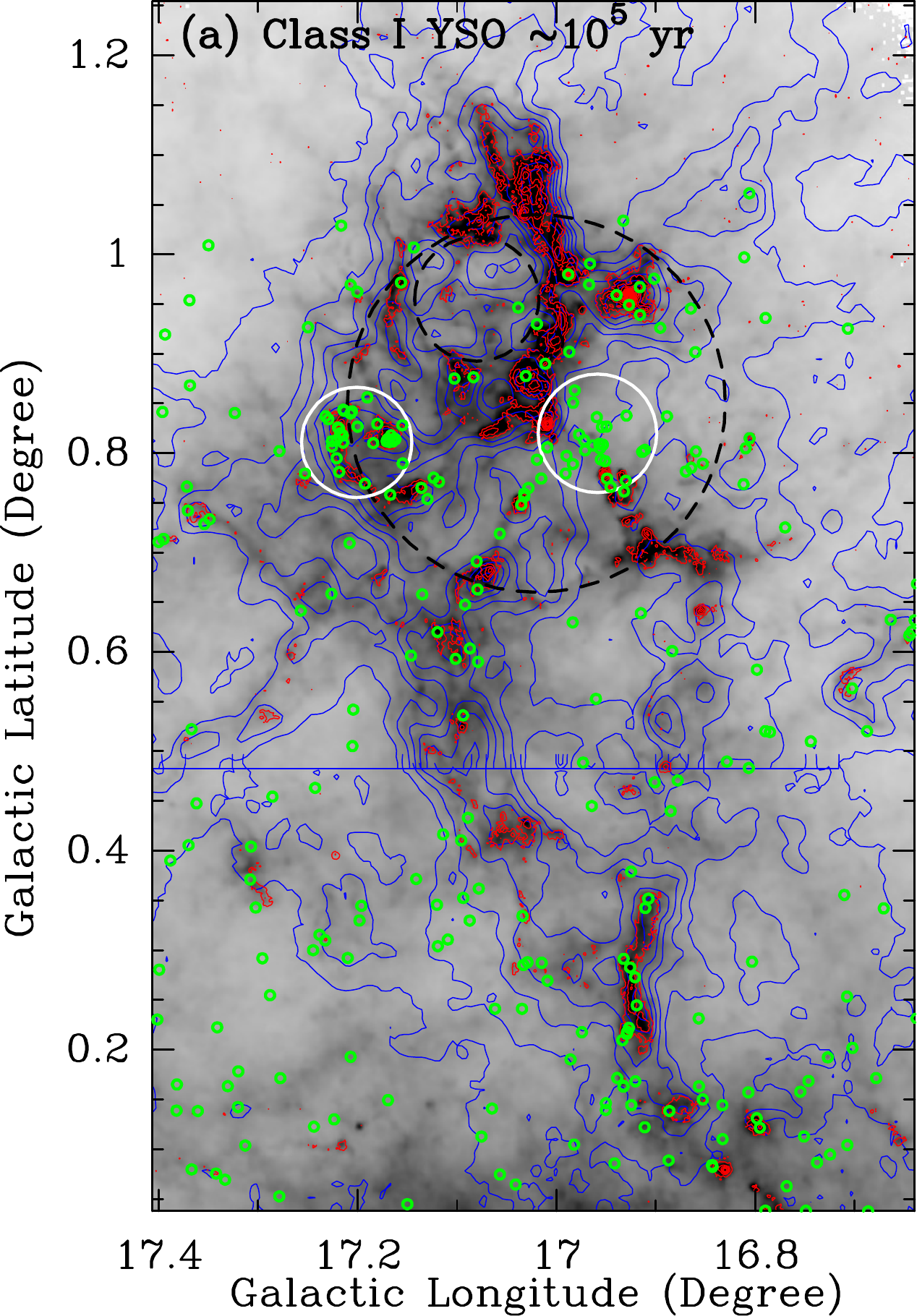}
\includegraphics[width = 0.45 \textwidth]{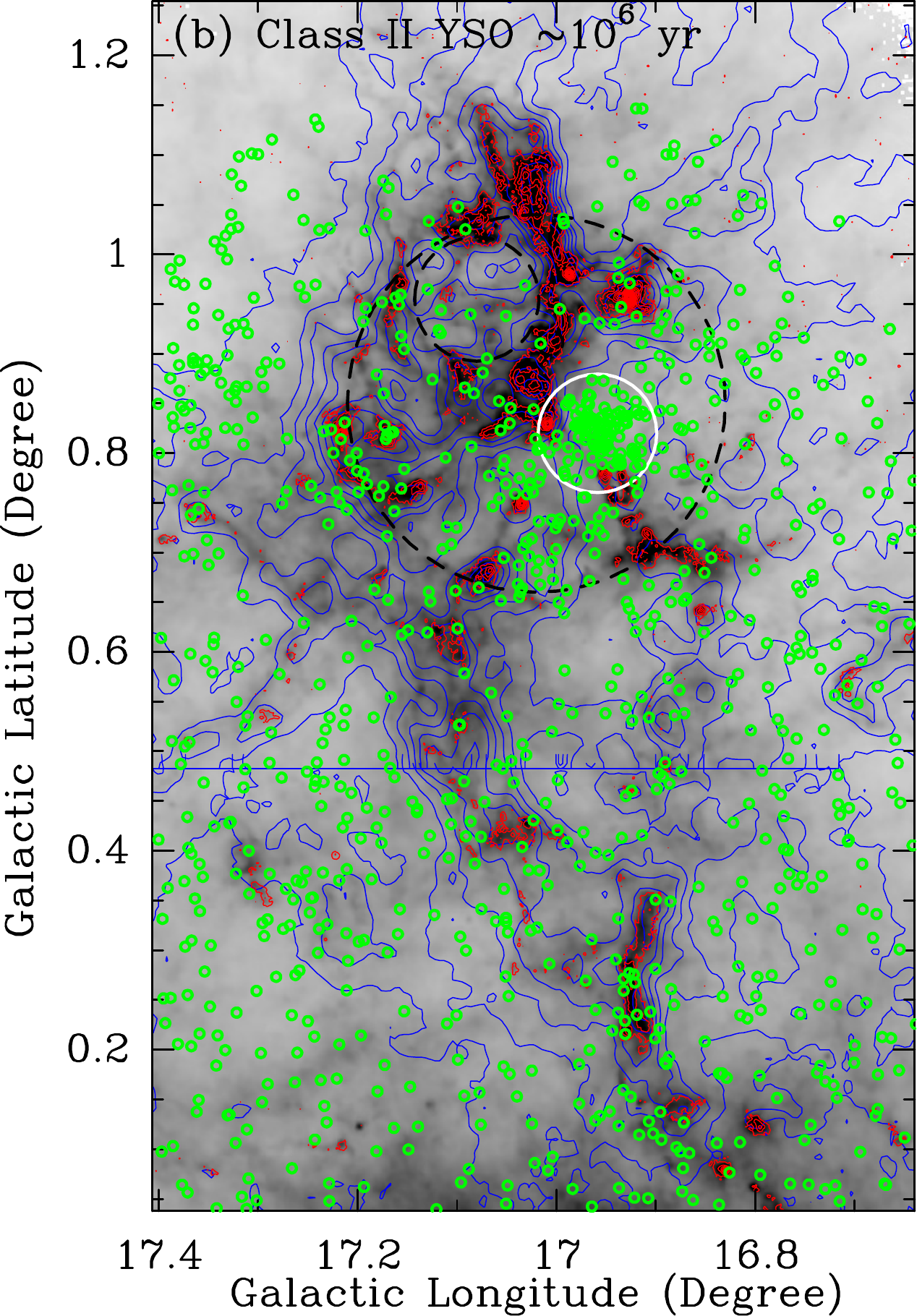}
\caption{The $^{13}$CO $J$=1-0 emission (blue contours) and 870 $\mu$m emission (red contours) are superimposed  on the {\it Herschel} 250 $\mu$m emission in grayscale. The green circles represent the identified class I and II YSOs. The white circles outline the regions of YSO accumulation. (a): Distribution of the Class I YSOs. (b): Distribution of Class II YSOs.}
\label{Fig:YSO}
\end{figure*}

\subsubsection{Excitation temperature, optical depth, and velocity dispersion }
The CO molecular gas associated with  M16 has three different velocity components. The dust cores only show the 2D projected structure. We therefore need to determine to which component each core is linked.  Within the effective radius of each core, we search for $^{12}$CO $J$=1-0, $^{13}$CO $J$=1-0, and C$^{18}$O $J$=1-0 spectra that are located at or near the peak positions of the cores. We use three-velocity Gaussian components to fit the spectra.  Considering  whether the peak position of each core is associated with those of $^{13}$CO $J$=1-0  and/or C$^{18}$O $J$=1-0 emissions, we lastly select a fitted component as the spectral parameters of the core. Compared with other components, we found that the strongest CO components  are often associated with the cores.  For part of the cores for which PMO CO data is not available, we use GRS $^{13}$CO $J$=1-0 and JCMT $^{12}$CO $J$=1-0 survey data to obtain the parameters. The fitted parameters are listed in Table 2, including brightness temperature ($T_{\rm mb}$), $FWHM_{\rm spec}$, and centroid velocity ($V_{\rm LSR}$).  For some cores, we cannot give the effective spectral value because of the weak signal ($\leq 3 \sigma$).  Since the cores are spatially unresolved by the CO data, the spectra used for each core also cover the surrounding matter. Hence, the fitted parameters may be underestimated for each core.

As $^{12}$CO  emission is considered to be optical thick, we use $^{12}$CO  to calculate $T_{\rm ex}$ via  following the equation  \citep{Garden}
\begin{equation}
\mathit{T_{\rm ex}}=\frac{5.53}{{\ln[1+5.53/(T_{\rm mb}+0.82)]}},
\end{equation}
where $T_{\rm mb}$ is the corrected main-beam brightness temperature of $^{12}$CO. The derived excitation temperature in these cores ranges from 9.0  to 37.8 K with a mean value of 22.5 K.  Figure \ref{Fig:W37}d shows the positional distribution of the cores with different temperatures. The cores with temperature $>$ 20 K are mainly located in the northern part of the large-scale filament, which is interacting with the M16 \HII region.  In addition, we assume that the excitation temperatures of $^{12}$CO and $^{13}$CO have the same values in both cores.  The optical depth ($\tau$) can be obtained by the following equation \citep{Garden}
\begin{equation}
\mathit{\tau(\rm ^{13}CO)}=-\ln[1-\frac{T_{\rm mb}}{5.29/[\rm exp(5.29/\it T_{\rm ex})\rm -1]-0.89}].
\end{equation}
The derived optical depth is 0.2-1.9 with a median value of 0.5,  indicating that the $^{13}$CO emission is optically thin in most of the cores.  Hence, we use the $^{13}$CO V$_{\rm LSR}$  to determine the distance of  each core. The distances to the cores are estimated using the Bayesian Distance Calculator 4 \citep{Reid2016}.  We obtain that the distance of these cores is about 1.85$\pm$0.2 kpc.  The distance to NGC6611 is estimated to be in the range 1.75--2.00 kpc \citep{Hillenbrand1993,Loktin2003,Guarcello2007,Wolff2007,Gvaramadze2008}. M16 is excited by numerous O and B stars within the open cluster NGC6611. By comparing the distances obtained for the cores,  we demonstrate that the cores are associated with the M16 \HII region.

The 1D thermal velocity dispersion in each core, $\sigma_{\upsilon}$,  is determined by 
\begin{equation}
\mathit{\sigma_{\upsilon}}=\sqrt{\sigma_{\rm Therm}^{2}+\sigma_{\rm NT}^{2}}
,\end{equation}
where $\sigma_{\rm Therm}$ and $\sigma_{\rm NT}$ are the thermal and nonthermal 1D velocity dispersions. For all the cores, $\sigma_{\rm NT}$ and $\sigma_{\rm Therm}$ can be obtained using, respectively,
\begin{equation}
\mathit{\sigma_{\rm NT}}=[\sigma_{\rm C^{18}O}^{2}-\frac{kT_{\rm kin}}{m_{\rm C^{18}O}\mu}]^{1/2}
,\end{equation}
\begin{equation}
\mathit{\sigma_{\rm Therm}}=\sqrt{\frac{kT_{\rm kin}}{m_{\rm H}\mu}} 
,\end{equation}
where $T_{\rm kin}$ is the kinetic temperature in the core. If the densities of the cores are high enough so that LTE conditions hold,  $T_{\rm ex}$ can be adopted as $T_{\rm kin}$ \citep{Feher2017}. For the dense ATLASGAL cores, here we take $T_{\rm ex}$ as $T_{\rm kin}$. Further,
 $\mu$=2.72 is the mean atomic weight, $m_{\rm H}$ is the mass of an H atom, $\sigma_{\rm C^{18}O}$ = ($\Delta V_{18}/\sqrt{\rm 8ln2}$) is the 1D velocity dispersion of C$^{18}$O $J$=1-0, and $m_{\rm C^{18}O}$ is the mass of C$^{18}$O $J$=1-0.  For the 51 cores, only 35 cores have C$^{18}$O $J$=1-0 emission above $3\sigma$. Their derived parameters are summarized in Table 3. The mean thermal 1D velocity dispersion of these cores ranges from 0.22 to 0.33 km s$^{-1}$ with a mean value of  0.27 km s$^{-1}$.  The nonthermal 1D velocity dispersion of these cores ranges from 0.34 to 1.34 km s$^{-1}$ with a mean value of 0.84 km s$^{-1}$.
The mean 1D velocity dispersion $\sigma_{\upsilon}$ ranges from 0.52 to 1.36 km s$^{-1}$ with a mean value of 0.87 km s$^{-1}$. Figure \ref{Fig:Tex} shows the distributions of excitation temperature, optical depth, and velocity dispersion; we obtain that the nonthermal motion is dominating for the velocity dispersion.

{\subsubsection{Mass, size,  and volume density}
We use the 870 $\mu$m flux of the cores to estimate their mass. Assuming optically thin emission, the mass is given by \citep{Hildebrand1983}
\begin{equation} \mathit{M_{\rm clump}}=\frac{S_{\nu}D^{2}}{\kappa_{\nu}B_{\nu}(T_{d})}
,\end{equation}
where $S_{\nu}$ and $D$  correspond to the flux density at the frequency $\nu$ and the distance to the cores. The ratio of gas to dust was adopted as 100 \citep{Schuller2009}, $\kappa_{\nu}$ is the dust opacity, which is adopted as 0.01 cm$^{2}$ g$^{-1}$  at 870 $\mu$m \citep{Ossenkopf1994}, and $B_{v}(T_{\rm d})$ is the Planck function for the dust temperature $T_{d}$ and frequency $\nu$.  The gas and dust temperatures are coupled if the gas densities are higher than 2$\times$10$^{4}$ cm$^{-3}$ \citep{Goldsmith2001, Galli2002}.  \citet{Csengeri2017} obtained that the mean gas density of the ATLASGAL clumps associated with GMCs is $\sim$3$\times$10$^{5}$ cm$^{-3}$. Hence, we take the excitation temperature as a dust temperature to estimate the  mass of these cores.

Assuming that the cores have roughly spherical shapes, the average volume density of each core was calculated as 
\begin{equation}
\mathit{n_{\rm H_{2}}}=\frac{M}{4/3\pi R_{\rm eff}^{3}\mu m_{\rm H}},
\end{equation}
where $R_{\rm eff}$ is the effective radius of each core, which is determined by 
\begin{equation}
\mathit{R_{\rm eff}}={D\sqrt{FWHM/2}},
\end{equation}
where $FWHM$ is the deconvolved size of the cores, respectively. The obtained $R_{\rm eff}$ are listed in column 3 of Table 3. Core radius ranges from 0.09 to 0.25 pc. The masses of these cores range from 11.9$\pm$1.8 to 357.3$\pm$53.6 M$_{ \odot}$ with a total mass of 2716 M$_{ \odot}$, while the average volume density ranges from 2.5$\times$10$^{4}$ to 5.9$\times$10$^{5}$ cm$^{-3}$. The typical uncertainties of the parameters largely originate from the uncertainties in flux estimation.  The final flux uncertainty for the compact cores should be lower than 15\% \citep{Schuller2009}, which can propagate to other parameters.

\subsection{Young Stellar Objects Associated with the Larger Filament}
Since the {\it Spitzer}-IRAC bands are highly sensitive to the emission from the circumstellar disks and  envelopes, the IRAC color-color diagrams are used to identify young stellar objects (YSOs) with infrared excess in star-forming regions and categorizing them according to evolutionary stages  \citep{Allen}. In order to study the YSO population within the large-scale filament,   we use the GLIMPSE I Spring'07 catalog  to search for YSOs. Based on the criteria of \citet{Allen}, we selected near-infrared sources with 3.6, 4.5, 5.8, and 8.0 $\mu$m detections from the catalogue.  All selected sources have  photometric uncertainties$<$0.2 mag in all four IRAC bands.  In the large-scale filament associated with the M16 \HII region,  we found 267 Class I YSOs and 886 Class II YSOs.  Class I YSOs are protostars with  circumstellar envelopes and a timescale of the order of $\sim10^{5}$ yr, while Class II YSOs are disk-dominated objects with a $\sim10^{6}$ yr \citep{Andre94}. The Class III YSOs are the pre-main sequence stars.  Figure \ref{Fig:YSO} shows the spatial distribution of these Class I and Class II YSOs. To investigate the positional relation of YSOs with the large-scale filament, we also overlaid the selected Class I and Class II YSOs on the $^{13}$CO $J=1-0$ emission (contours) and 250 $\mu$m emission (gray scale). Here the integrated-velocity  range (16--24 km s$^{-1}$) of $^{13}$CO $J=1-0$ covers all the components. Class I sources in Figure \ref{Fig:YSO}a are found to be mostly concentrated in two regions, which are marked by two white dashed circles. One region is associated with the NGC6611 cluster  responsible for the ionization of M16, the other region is situated  at the edge of M16. Although most of the Class II YSOs in Figure \ref{Fig:YSO}b appear to be dispersedly distributed across the selected region,  there are still some Class II YSOs concentrated in the position of the NGC6611 cluster. Since the whole region has many Class III YSOs, we do not show  the distribution of these YSOs in Figure \ref{Fig:YSO} to avoid confusion.

\section{Discussion}
\label{sect:discu}
\subsection{Gas structure associated with M16}
The CO molecular gas and cool dust emissions are mainly located in the northeastern part of the M16 \HII region. The expansion of  M16 is likely to be blocked by the molecular gas along this direction. The western CO and dust emission of M16 are very weak, allowing the expansion of M16 to break the molecular gas and create a  cavity.  On large scale, the $^{13}$CO $J$=1-0 emission  consists of three velocity components.  Each component shows a filamentary structure extended along the north-south direction. The spatial overlap of these three components along the line of sight suggests that the large-scale filament has three layers. From component 1 to component 3, we find that several pillars, revealed by their $^{13}$CO $J$=1-0 emission, are distributed over the edges of the entire M16 region.  The northeastern part of the large-scale filament may be disrupted into some small IRDCs because of the expansion of the M16 \HII region, which is similar to  the case of the filamentary IRDC G34.43+0.24 \citep{Xu2016}.  The presence of several smaller IRDCs and pillars suggests that M16 is interacting with the large-scale filament and has reshaped its structure. Both the large-scale filament associated with M16 and the filamentary IRDC G34.43+0.24 show the 8 $\mu$m IR dark and bright parts. The bright parts of the IRDC are associated with \HII regions and IR bubbles. Hence, the associated \HII regions may illuminate the dark parts of the large-scale filament.

Moreover, to investigate the  dynamic structure of the molecular gas surrounding the N19 bubble, we  made a position--velocity (PV) diagram along the northwestern direction towards the NGC6611 cluster. The cutting direction goes through N19. In Figure \ref{Fig:PV}a, the two red vertical lines mark the edges of N19. The diameter of the bubble is obtained from Figure \ref{Fig:CO-8UM}. The green arrows are assumed to indicate the direction of the shock from the M16 \HII region, while the blue dashed line  marks the compressed surface of the molecular cloud. From Figure \ref{Fig:PV}a, we can see that the  M16 \HII region is probably interacting with three molecular layers.  Furthermore, we also observe the CO emission that  peaks at 16 km s$^{-1}$ towards N19, which is likely to be the front or back side of the bubble if the bubble is a sphere.  A molecular pillar marked by a blue arrow at the edge of N19 might have been created by the expansion of N19. In Figure \ref{Fig:CO-8UM}, the 870 $\mu$m emission delineates the  NS filament. To explore the structure of this NS filament, we use two directions to make the PV diagram along the NS filament constructed from the $^{13}$CO $J$=1-0  emission. These directions are shown in Figure \ref{Fig:CO-8UM}.  As seen in Figure \ref{Fig:PV}, the northern part of the NS filament shows a single component, suggesting a single coherent object which is associated with component 1 in velocity. The remaining part contains three different velocity components. Hence, we conclude that the whole NS filament seen in the {\it Herschel} data  and on the ATLASGAL 870 $\mu$m emission is  the projection effect from the molecular gas of the different velocity components.

\begin{figure*}
\includegraphics[width = 0.51 \textwidth]{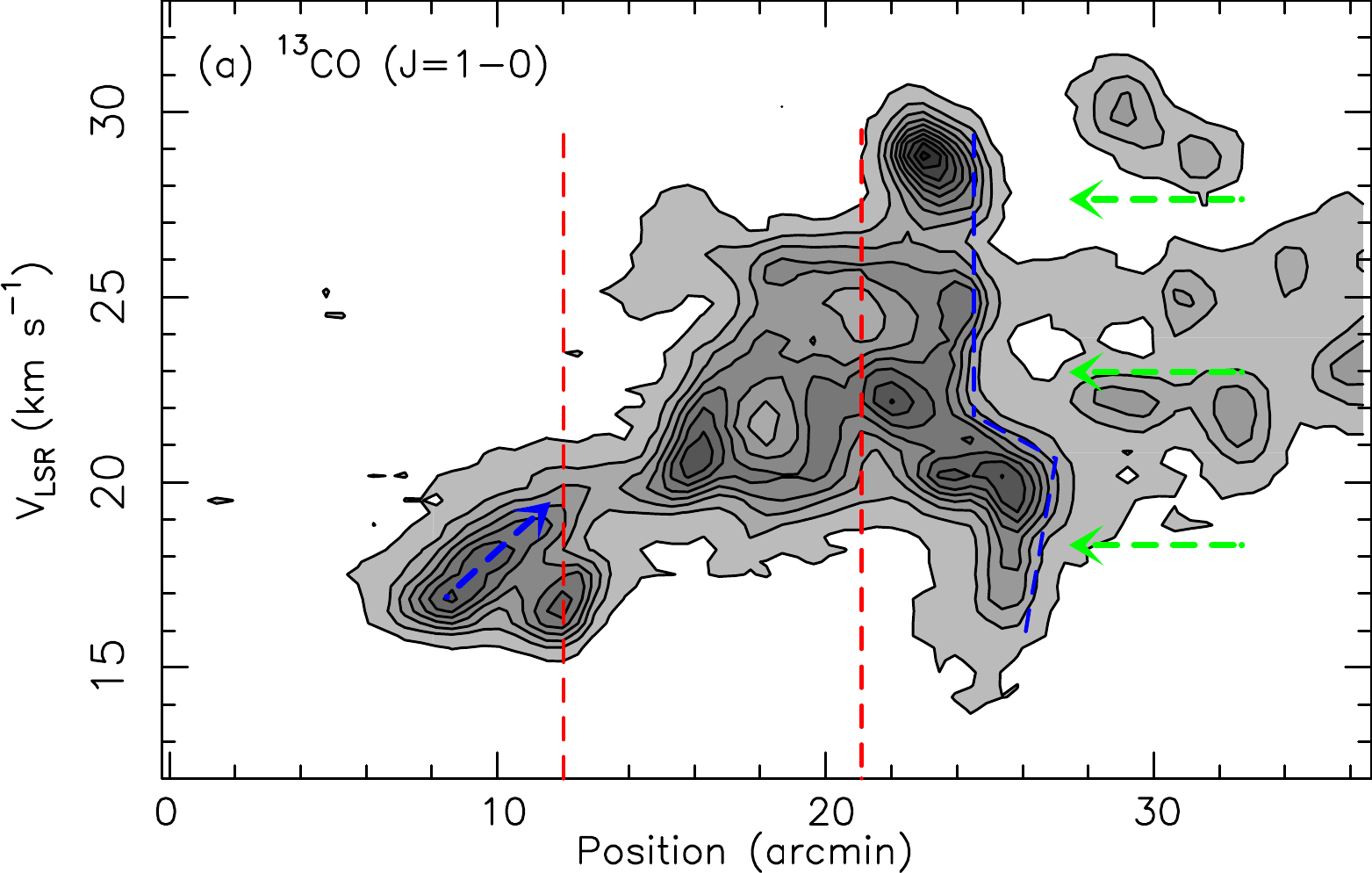}
\includegraphics[width = 0.49 \textwidth]{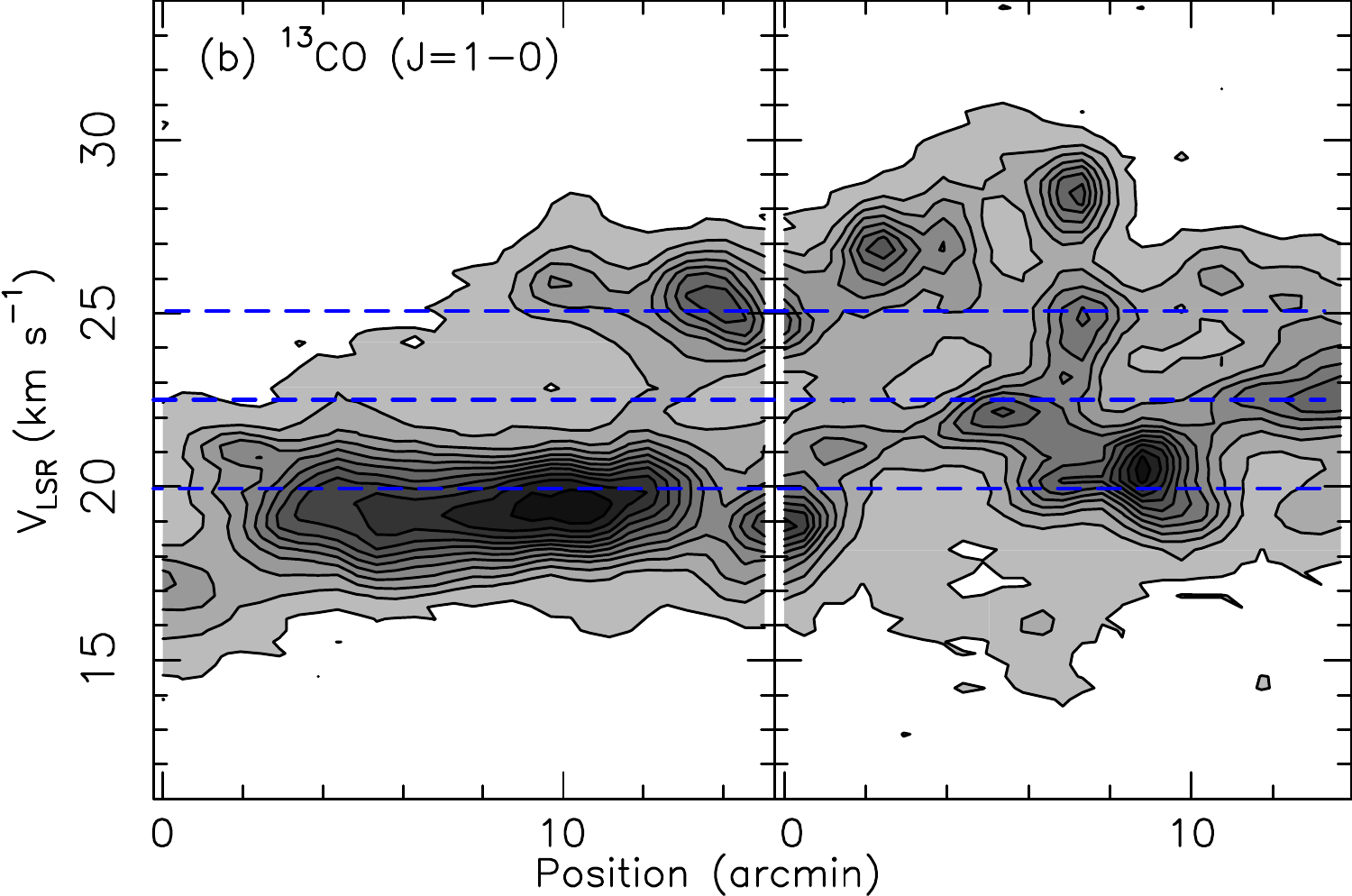}
\caption{(a): Position-velocity diagram of the $^{13}$CO $J$=1-0 emission along the northwest of the N19 bubble (see the green dashed lines in Figure \ref{Fig:CO-8UM} a panel). The red dashed lines mark the edges of the N19 bubble. The green arrows show the direction of the interaction between the  
HII region of M16 and the molecular gas. The blue arrow shows the high-velocity gas, while blue dashed lines indicate the compressed positions. (b): Position--velocity diagram of the $^{13}$CO $J$=1-0 emission along the NS filament (see the green dashed lines in Figure \ref{Fig:CO-8UM} a panel). The blue dashed lines shows the three velocity components, which are mentioned in Sect. 3.2.  We refer to the discussion about the PV diagrams in Section 4.1. }
\label{Fig:PV}
\end{figure*}

\begin{figure*}
\centering
\includegraphics[width = 0.55 \textwidth]{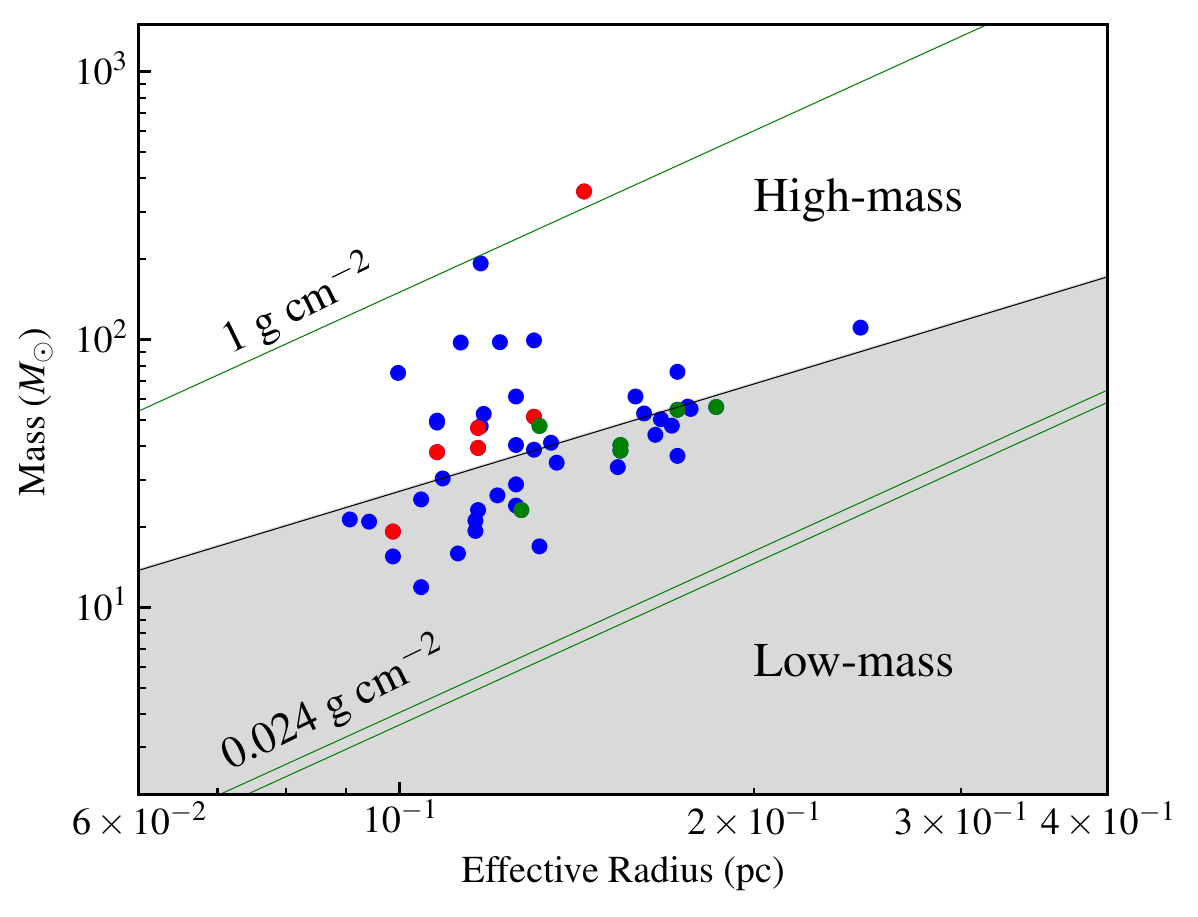}
\vspace{-2mm}
\caption{ Mass--radius distributions for the ATLASGAL 870 $\mu$m cores. The green dots show the six cores in the small filament, while the red dots show the six cores in the large clump. 
The surface density of 0.024 g cm$^{-2}$ shown in green lines gives the average surface density thresholds for efficient star formation, which are derived by \citet{Lada2010} and \citet{Heiderman2010}, respectively.  A core with a mass surface density of $>$ 1 g cm$^{-2}$ (shown as a green line) can avoid further fragmentation and form massive stars \citet{Krumholz2008}. We refer to the discussion about the  lines in the Sect. 4.2.} 
\label{Fig:mass}
\end{figure*}

\subsection{Cores properties}
In the large-scale filament, we found 51 dust cores. The excitation temperatures in these cores range from 9.0 to 37.8 K with a mean value of 22.5 K. Generally, dust is heated by UV and IR emission, while gas is heated only by UV through photo-electric heating \citep{Hollenbach1997}.  Previous observations indicated that  cold dark cores may have a typical excitation temperature of $\sim$ 10.0 K \citep{Du2008,Meng2013,Liu2014}.
Compared to the cold dark cores, the dust cores in the large-scale filament are likely to be heated by the UV emission from the M16 \HII region.  \citet{Guzman2015} studied about 3000 molecular clumps from ATLASGAL data at 870 $\mu$m. They obtained a mean dust temperature of 16.8 K for the quiescent clumps, 18.6 K for protostellar clumps, and 23.7 K and 28.1 K for clumps associated with \HII and PDR, respectively.  Through visual inspection of {\it Spitzer} images at 3.6, 4.5, 8.0, and 24 $\mu$m, \citet{Foster2011} developed a criterion for classification of core evolution. Specifically, cores which are dark at GLIMPSE wavelengths (3.6-8 $\mu$m) are identified as quiescent, cores with a MIPSGAL 24 $\mu$m point source are identified as protostellar, and cores with extended 8 $\mu$m flux are identified as \HII regions. Based on the criterion of \citet{Foster2011}, we classify the 51 cores in the large-scale filament into three evolutionary stages. Column 12 in  Table 3 gives the classification of the cores, as quiescent (20 cores), protostellar (3 cores), \hii/PDR (28 cores). Because  some cores are distributed close to the  M16 \HII region, the \hii/PDR classification carries significant uncertainty.  The excitation temperatures of the 20 identified quiescent cores in the  large-scale filament range from 14.2 to 30.8~K with a mean value of 22.2 K.   Some of the quiescent cores are also called IR dark cores  in \citet{Guzman2015}. Assuming that dust and gas temperatures are similar (see Sect. 3.3.2), the mean value (22.2 K) of the temperature for quiescent cores in the  large-scale filament is higher than that (16.8 K) of \citet{Guzman2015}, suggesting that the cores in the large-scale filament are heated by the radiation of the M16 \HII region, not by an internal heating due to sources in the cores. At a temperature $<$ 20 K, the molecular gas is cold \citep{Guzman2015, Egan1998, Carey1998}. In the large-scale filament,  there are 34 cores whose excitation temperature is higher than 20.0 K, meaning that at least 67$\%$ of the cores have been heated. From Figure \ref{Fig:W37}d, we also see that the majority of the heated cores is spatially close to the M16 HII region. 

To determine whether these cores have sufficient mass to form massive stars, we can use the mass--size relation given by \citet{Kauffmann2010}. The radii of the cores range from 0.09  to 0.25 pc, while the masses range from 12 to 357  {\it M$_{ \odot}$}.  Figure \ref{Fig:mass} presents a mass-versus-radius  plot for all the cores.  The surface density of  0.024 g cm$^{-2}$ gives the average surface density thresholds for efficient star formation. The thresholds, shown as green lines in Figure  \ref{Fig:mass},  are derived by \citet{Lada2010} and \citet{Heiderman2010}, respectively.  As can be seen from Figure  \ref{Fig:mass}, 51 cores (100\%)  are located above the lower surface density limit of  0.024 g cm$^{-2}$.  \citet{Krumholz2008} suggested that the cores with a mass surface density of $>$ 1 g cm$^{-2}$ can avoid further fragmentation and form massive stars. Moreover,  \citet{Kauffmann2010} investigated the mass-radius relationship of cores and found the empirical relationship $M(r)\geq870M_{\odot}(r/\rm pc)^{1.33}$ as the threshold to determine whether cores can potentially form massive stars. We note  that  when deriving  their  relationship, \citet{Kauffmann2010}  reduced  the  dust opacities  of  \citet{Ossenkopf1994}  by  a  factor  of  1.5. Because this correction has not been applied here, we have rescaled the relationship $M(r)\geq580M_{\odot}(r/\rm pc)^{1.33}$ , which is also given  by  \citet{Urquhart2013}.  We find that about 45\% of all the identified cores are above the rescaled threshold, indicating that these cores are dense and massive enough to potentially form massive stars. In the bubble-shaped nebula Gum 31, 37\% of the 870 $\mu$m clumps lie in or above the massive-star formation threshold \citep{Duronea2015}. We suggest that HII regions can help in creating a larger number of massive clumps and cores, but more examples are needed to make a firm conclusion.

\subsection{Core formation}
In a filament, core formation may be regulated by the interplay between gravity, turbulence, and magnetic field  \citep{Li2015}. Several pillars are found in the large-scale filament associated with M16.  Particularly, there are some cores that are situated in the heads of the pillars. With the expansion of an \HII region, \citet{Schneider2012} suggested that when the ionized  gas pressure  dominates the ram pressure of the initial turbulence in the molecular cloud, this leads to the formation of the cores and pillars}. This is a strong indication that the pillars in M16 are produced by the interplay of pre-existing turbulent structures and ionizing radiation \citep{Gritschneder2010}. Here at least 67$\%$ of the cores in the large-scale filament have been heated by the  M16 \HII region. However, compared with the thermal motion, we suggest that the nonthermal motion may dominate the velocity dispersion in the cores of the large-scale filament. The nonthermal motions in the cores
are generally interpreted as being due to supersonic turbulence \citep[e.g.,][]{Zuckerman1974,McKee2007}, which is responsible for the core formation in the large-scale filament. Using C$^{18}$O $J$=1-0, the nonthermal 1D velocity dispersion $\sigma_{\rm NT}$ is 0.27 km s$^{-1}$ for the Taurus molecular cloud (TMC), and 0.35 km s$^{-1}$ for the  California molecular cloud (CMC) \citep{Meng2013}. The TMC is a typical site of low-mass star formation. The CMC is  in an early state of evolution and has not achieved the internal physical conditions to promote more active star formation \citep{Lada2009}. Hence, both TMC and CMC  have less star-forming activity  compared to M16. The mean $\sigma_{\rm NT}$ of these cores is 0.84 km s$^{-1}$ for  C$^{18}$O $J$=1-0 in the large-scale filament associated with the M16 \HII region. Compared with the TMC and CMC, the higher $\sigma_{\rm NT}$ in the large-scale filament may be created by the feedback from the M16 \HII region. As the pillars are seen adjacent to M16, the \HII region may give rise to the strong turbulence in  the large-scale filament.

In molecular clouds, turbulence has been shown to dissipate quickly \citep{Stone1998,Mac1999,Orkisz2017} if the external driving is stopped, resulting in the need to continuously drive turbulence via stellar feedback \citep{Ostriker2010,Offner2018}. Here we can estimate the turbulent energy of the large-scale filament, which is given by
\begin{equation}
\mathit{E_{\rm turb}}=\frac{1}{2}M\sigma^{2}_{3d},
\end{equation}
where $\sigma_{3d}$$\approx$$\sqrt{3}$$\sigma_{\rm NT}$, which is the 3D turbulent velocity dispersion. In the large-scale filament, we found the mean $\sigma_{\rm NT}$ to be 0.84 km $\rm s^{-1}$, which is obtained from Sect. 3.3. The mass of the filament can be obtained by

 $M$=$\mu m_{\rm H} N_{\rm H_{2}}S$, 

where $N_{\rm H_{2}}$ is the column density, and $S$ is the area of the large-scale filament, which can be determined by the {\it {\it Herschel}}  250 $\mu$m data in Figure \ref{Fig:W37}. Using the {\it {\it Herschel}} data and spectral energy distribution (SED) fitting, \citet{Hill2012} obtained that the column density of the dense large-scale filament associated with M16 is (3-7)$\times10^{22}$ cm$^{-2}$.  We derive a mass of (4.3-9.9)$\times10^{5}$ $M_{ \odot}$ for the large-scale filament and estimate the turbulent energy of the large-scale filament to be about (1.8-4.2)$\times10^{49}$erg. The formation of the M16 shell may be attributed to the expansion of  the M16 \HII region that is ionized by a few O and B stars. \citet{Sofue1986} obtained that the total kinetic energy ejected from these O and B stars is evaluated as about 7 $\times10^{50}$ erg, which is  one order of magnitude higher than the obtained turbulent energy in the large-scale filament.  For an \HII region, \citet{Freyer2003} and \citet{Xu2018} indicated that the ionization energy is one order of magnitude higher than the kinetic energy and thermal energy. Hence, the energies from the M16 \HII region  can help to maintain the strong turbulence in the large-scale filament via energy injection.

\begin{figure*}
\centering
\includegraphics[width = 0.51 \textwidth]{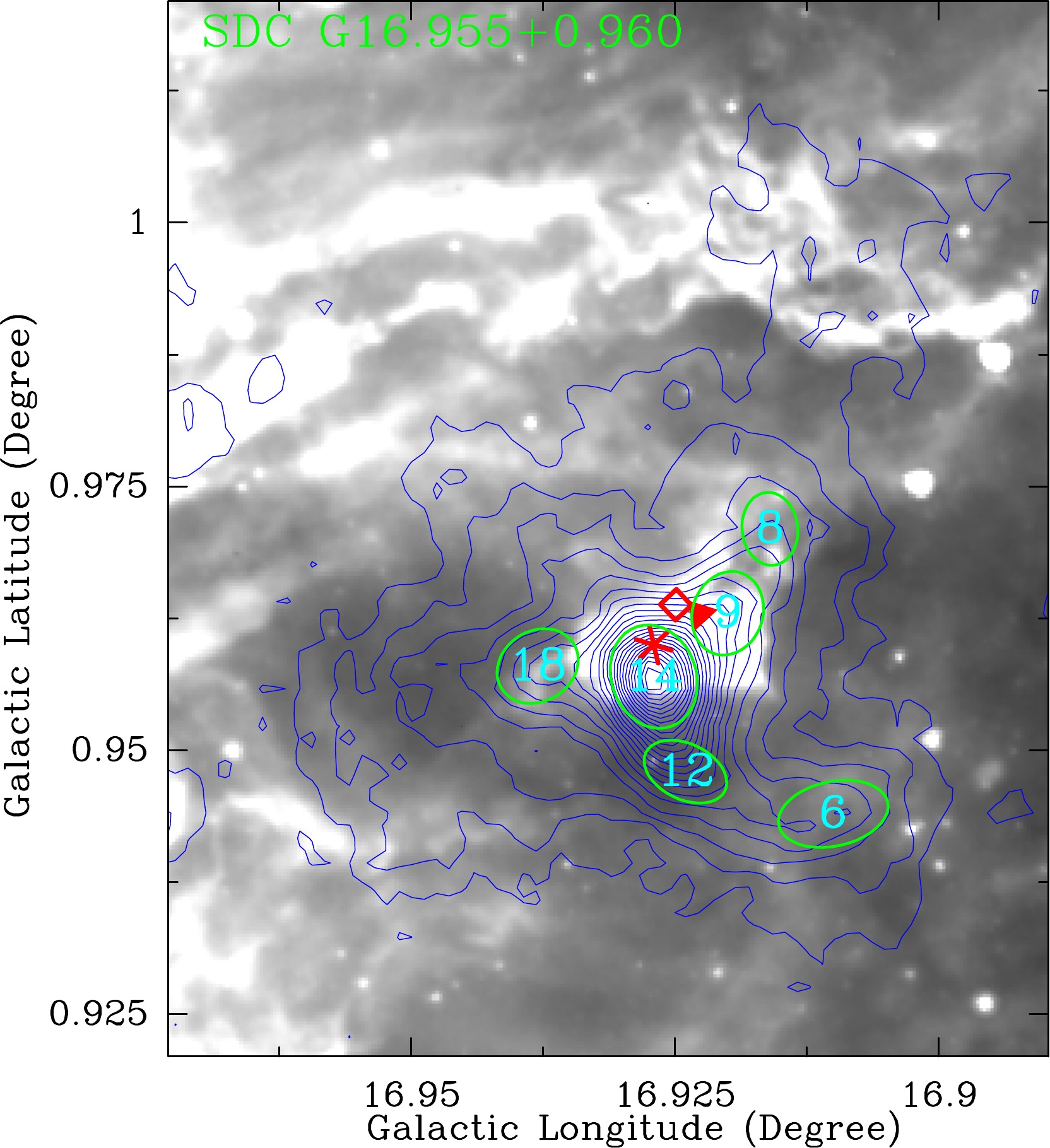}
\includegraphics[width = 0.47 \textwidth]{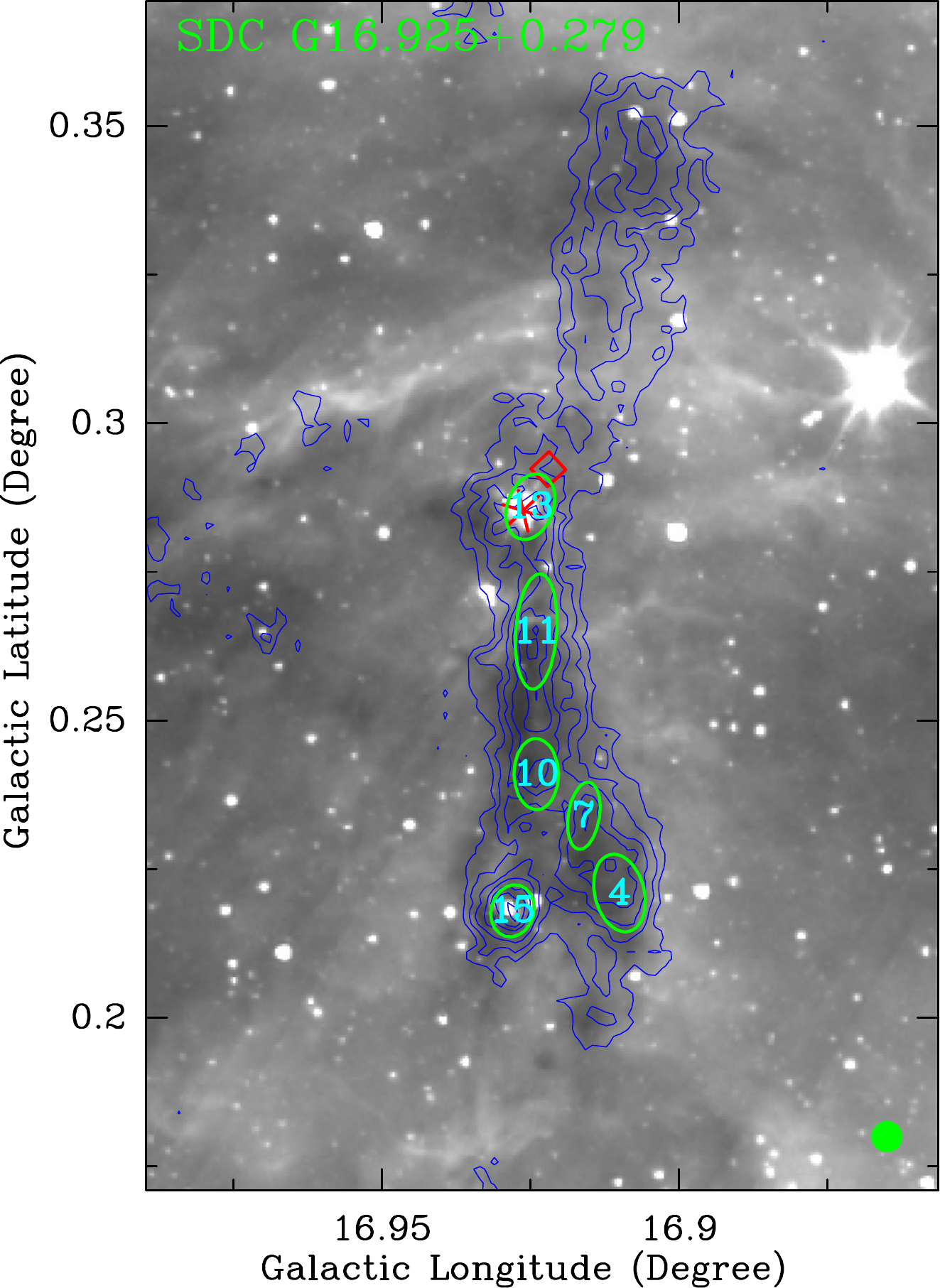}
\caption{The 870 $\mu$m emission (blue contours) superimposed on the 8 $\mu$m emission (gray). The two regions are as also shown in the two white dashed boxes in Figure \ref{Fig:W37}.  The blue contour levels start from 3 $\rm\sigma$ (0.15 Jy/beam) and rise in steps of 2 $\rm\sigma$. The green ellipses represent the 870 $\mu$m cores. The red stars mark the positions of the MSX6C G016.9270+00.9599 and MSX6C G016.9261+00.2854 \HII regions \citep{Urquhart2011}. The red squares indicate IRAS sources, while the red triangle  shows the water maser \citep{Braz1983}. The ATLASGAL beam (19$^{\prime\prime}$) is shown as the green filled circle in the bottom right-hand corner.}
\label{Fig:W37-small}
\end{figure*}

\subsection{Turbulent fragmentation}
In the large-scale filament, there are six different compact cores in a large clump and in a small filament (southern filament),  as shown in Figures \ref{Fig:W37-small}a and \ref{Fig:W37-small}b, respectively. The large clump is associated with the dark IR cloud SDC G16.955+0.960, while the  southern filament is consistent with SDC G16.925+0.279 \citep{Peretto2009}. Because both the clump and filament are also associated with an \HII region  \citep{Urquhart2011} and an IRAS source, they are massive star-forming regions. In the large clump, the mean excitation temperature and $\sigma_{v}$ of the  six cores are 19.9 K and 0.93 km s$^{-1}$,  respectively.  Adopting the mean excitation temperature as the dust temperature (see Sect. 3.3.2), we obtained a mass   of 2569 $M_{ \odot}$ for the large clump. Figure \ref{Fig:mass} shows that five of the six cores (red dots) in the large clump are dense and massive enough to potentially form massive stars. Particularly, the mass surface density of  core 14 is higher than 1 g cm$^{-2}$. According to \citet{Krumholz2008}, this means that because of the prevention through radiative feedback, this core will not further fragment into low-mass cores, thus allowing high-mass star formation. Moreover, core 14 is also associated with the  MSX6C G016.9270+00.9599 \HII region in space \citep{Urquhart2011}, as shown in Figure \ref{Fig:W37-small}a, therefore further supporting the results of \citet{Krumholz2008}. Assuming that the clump is governed by Jeans instability, we estimate the Jeans mass, which is given as  $M_{\rm J} = 0.877(T/10 \rm K)^{3/2}$$(n/10^5 \rm cm^{-3})^{-1/2}$$M_{ \odot}$  \citep{Wang2014}. From Eq. 7, we derive the number density $n=9450$ $\rm cm^{-3}$. Taking the mean  excitation temperature (19.9 K) of the six cores as that of the clump, we determine $M_{\rm J} \sim 8.0 M_{ \odot}$. The masses of the six cores in the clump range from 19 to 357 $M_{ \odot}$, which is larger than the Jeans mass (8.0 $M_{ \odot}$), indicating that these cores may have formed in the clump through the turbulent core model \citep{McKee2002}.

For the southern filament, the mean excitation temperature of the six cores is 19.9 K, which is the same as that in the clump. From Figure \ref{Fig:mass}, we  see that only one core is massive enough to form massive stars in the southern filament, which is different from that in the clump. If the turbulence dominates  in the southern filament,  the critical linear mass density can be estimated as $(M/l)_{\rm crit}$=84($\rm \Delta V)^{2}$ $M_{ \odot}$ pc$^{-1}$ \citep{Jackson2010}, where $\rm \Delta V$ is the linear width in units of km s$^{-1}$.  We did not observe the   southern filament in C$^{18}$O. The mean $\rm \Delta V$ of $^{13}$CO in the  southern filament is 2.4 km s$^{-1}$, while the mean C$^{18}$O $\rm \Delta V$ of the large-scale filament is also 2.4 km s$^{-1}$.  Adopting a mean $\rm \Delta V$  of 2.4 km s$^{-1}$ for the  southern filament, we obtain  $(M/l)_{\rm crit} \sim$ 485.5 $M_{ \odot}$ pc$^{-1}$. From Figure \ref{Fig:W37-small}, we measure the length of the   southern filament to be 5.6 pc. Using Eq. 6,  the derived mass of  the   southern filament is 1613 $M_{ \odot}$. Using the mass and length, we derive a linear mass density of $M/l$= 289.6  $M_{ \odot}$ pc$^{-1}$, which is roughly consistent with ($M/l$)$_{\rm crit}$.  Considering the uncertainties of the FWHM, the turbulent motion  may help to stabilize  the  southern filament against radial collapse \citep{Jackson2010,Beuther2015}.

In addition, we can calculate a core formation efficiency (CFE) for the compact clump and the  southern filament, which is given by $M_{\rm core}$/$M_{\rm clump}$. The total mass of the cores in the clump is 552$\pm$26 $M_{ \odot}$, while this is 260$\pm$8 $M_{ \odot}$ in the southern filament. We obtain that the CFE for the clump is 22$\pm$3\%, while this value is 16$\pm$2$\%$ in the southern filament. The southern filament is likely to be compressed by the G17.037+0.320 \HII region  \citep{Anderson2014}, as shown in Figure \ref{Fig:W37}. Compared with the southern filament, the clump is closer to the M16 \HII region. The strong feedback from M16 may create the higher CFE. The CFE in the GMC of the Milky Way is about 11.0\% \citep{Battisti2014}. In the W3 GMC, the CFE in the filament compressed by the \HII region is in the range 26-37\%, while this value is 5-13\% in the diffuse region  \citep{Moore2007}. \citet{Eden2013} also gave a CFE in the region associated with  \HII regions of about  40\%. The obtained clump formation efficiency in the filament G47.06+0.26 associated with bubbles/\HII regions is $\sim$15\% \citep{Xu2018}. From the above analysis, we conclude that the molecular clouds associated with \HII regions have a higher CFE.  The higher CFE may be created by the ionization feedback  from the \HII regions.

\subsection{The NGC6611 cluster and YSO formation }
 We have shown that the M16 \HII region  interacts with a large-scale filament with three layers. M16 is ionized by numerous O and B stars within the open cluster NGC6611. Because the large-scale filament is associated with some IRDCs, as shown in the pink dashed lines in Figure \ref{Fig:8um-90cm},  the NGC6611 cluster  may form in a dark IR filament.  The age of the YSOs in NGC6611 is estimated to be 0.25--3 Myr \citep{Hillenbrand1993}.  We also use the {\it Spitzer}-IRAC data  to identify Class I  and Class II YSOs. Class I YSOs have  an age of  $\sim10^{5}$ yr, while this is $\sim10^{6}$ yr for Class II YSOs  \citep{Andre94}. The selected YSOs (Class I and Class II) are also concentrated in NGC6611. Furthermore, \citet{Hillenbrand1993} found that the stars of 3 $M_{ \odot}$ $<$ mass $<$ 8 $M_{ \odot}$ have ages ranging from 0.25 to at least 1 Myr, while the mean age of the stars  with a mass of $>$ 9  $M_{ \odot}$ is 2$\pm$1 Myr.  The stars with the ages from 0.25 to 1 Myr may belong to Class I and Class II YSOs. From the ages and masses of the cluster NGC6611 members, the massive stars may form before low-mass stars in a cluster. 
 
Additionally, a high density of Class I YSOs is also found to be located at the edges of the M16 \HII region. Since some of the selected Class I and Class II YSOs are clustered, it is unlikely that they are all simply foreground and background stars. It is more likely that most of the clustered Class I and Class II YSOs are physically associated with the large-scale filament. Through the Hertzprung-Russell (HR) diagram, \citet{Hillenbrand1993} obtained that the age of NGC6611 is likely to be about 2$\pm$1 Myr. Because M16 is excited by the O and B stars within the open cluster NGC6611, we assume that the dynamical age of M16 is the same as that of NGC6611.  Comparing the dynamical age of the M16 \HII region with those of the Class I YSOs, we conclude that the formation of these  Class I YSOs might have been triggered by the \HII region M16.  Moreover, the IR bubble N19 and RCW165 are  located in the dense region of the large-scale filament, and  interact with the filament. The ionized stars of N19 and RCW165 may be second-generation massive stars, whose formation was triggered by the expansion of the M16 \HII region.

\section{Conclusions}
\label{sect:summary}
Using the PMO molecular $^{12}$CO $J$=1-0, $^{13}$CO $J$=1-0, and C$^{18}$O $J$=1-0 data  combined with IR and radio archival  data, we present a comprehensive large-scale picture of the gas and dust towards the M16 \HII region. The main findings are summarized as follows.

1. The \HII region M16 shows an irregular ionized-gas cavity, which is enclosed by the cool dust traced by the 250 $\mu$m emission, and PAH emission traced by the 8 $\mu$m emission. From CO data, we  observe a large-scale filament with three main velocity components, whose peak velocities are 20.0, 22.5, and 25.0 km s$^{-1}$.  These three components overlap with each other, both in velocity and space, suggesting that the  large-scale is made of three layers.  Because the large-scale filament is associated with some IRDCs, the NGC6611 cluster may have formed in a dark IR filament. The presence of pillars associated with each velocity component indicates that the M16 \HII region is interacting with the large-scale filament and has reshaped the structure of the surrounding gas. 

2. In the whole large-scale filament, we find 51 dust cores from the ATLASGAL catalog and classify
the cores into three evolutionary stages, as quiescent (20 cores), protostellar (3 cores), and \hii/PDR (28 cores). We find that  45\% of all the identified cores are dense and massive enough to potentially form massive stars. The excitation temperature in these cores of the filament  range from 9.0 to 37.8 K with a mean value of 22.5 K. The mean excitation temperature of the identified quiescent cores is 22.2 K. The mean temperature for the quiescent cores suggests that the cores are externally heated by the M16 \HII region  and not internally due to sources in the cores. If the temperature of the heated cores is $>$20 K, here at least 67$\%$ of the cores have been heated. The majority of the heated cores is spatially close to the M16 \HII region.
 
3.  Compared with the thermal motion, the turbulence created by the nonthermal motion leads to the formation of the cores. Compared with the TMC and CMC, the higher nonthermal velocity dispersions in the large-scale filament may be created by the M16 \HII region.  Compared with the large-scale turbulent energy (1.8-4.2$\times10^{49}$erg), the  energies of the M16 \HII region can help to maintain the strong turbulence by injecting energy. 
A large clump and a southern filament contain six compact cores. The clump and the filament have been formed through the gas turbulence. 

4. In the large-scale filament, we find that the CFE for the clump is  22$\pm$3\%, while this value is 16$\pm$2$\%$ in the southern filament. The higher CFE may be created by feedback from the nearby \HII regions. 

5.  Comparing the dynamical age of the M16 \HII region with the Class I YSOs located at its edges, we suggest that the formation of these YSOs may have been triggered by the expansion of the M16 \HII region. The ionized stars of N19 and RCW165 may also be second-generation massive stars.

\begin{acknowledgements}
We thank the referee for insightful comments which improved the clarity of this manuscript.
We thank the Key Laboratory for Radio Astronomy, CAS, for partly supporting the telescope operation. This work  has made use of data from the Spitzer Space Telescope, which is operated by the Jet Propulsion Laboratory, California Institute of Technology under a contract with NASA. The ATLASGAL project is a collaboration between the Max-Planck-Gesellschaft, the European Southern Observatory (ESO) and the Universidad de Chile. This includes projects E-181.C-0885, E-078.F-9040(A), M-079.C-9501(A), M-081.C-9501(A) plus Chilean data. This work was supported by the National Natural Science Foundation of China (Grant Nos. 11673066, 11703040, and 11847309), the Youth Innovation Promotion Association of CAS,
the young researcher grant of national astronomical observatories, Chinese academy of sciences, and also supported by the Open Project Program of the Key Laboratory of FAST, NAOC, Chinese Academy of Sciences. AZ thanks the support of the Institut Universitaire de France.
\end{acknowledgements}

\bibliographystyle{aa}
\bibliography{references}

\clearpage
\begin{table*}
\renewcommand\arraystretch{0.9}
\begin{center}
\tabcolsep 3mm\caption{Parameters of the selected dust cores \citep{Contreras2013}. The columns are as follows: (1) the source ID; (2)-(3) the positions in galactic coordinates; (4)-(5) the beam-convolved major  and minor axes in arcseconds; (6) the position angle of the fitted Gaussian measured from north to east; (7) the average FWHM source size (beam-convolved); (8)-(9) the peak flux and the integrated flux.}
\def\temptablewidth{8\textwidth}%
\vspace{-2mm}
\begin{tabular}{llcccccccccccr}
\hline\hline\noalign{\smallskip}
ID    & l & b & $\theta_{\rm maj}$ & $\theta_{\rm min}$ & PA & $FWHM$ & F$_{\nu}$ & S$_{\nu}$ \\
   & (deg)   & (deg)   &  (arcsec)     &  (arcsec)  & ($^{\circ}$)  & (arcsec)  & (Jy/beam) & (Jy)  \\
  \hline\noalign{\smallskip}       
 1 &  16.804 & 0.814 & 23 & 19 & 80 & 21  & 0.48  & 0.56 \\
 2 &  16.854 & 0.642 & 26 & 23 & 33 & 24  & 0.97  & 1.56  \\
 3 & 16.890 & 0.484 & 28 & 25 & 79 & 26 & 0.55 & 1.04 \\
 4 & 16.910 & 0.221 & 48 & 30 & 104 & 38 & 0.4 & 1.59 \\
 5 &  16.908 & 0.719 & 32 & 19 &126 & 25  & 0.65  & 1.08  \\ 
 6 &  16.910 & 0.944 & 38 & 22 & 12 & 29  & 0.64  & 1.41  \\
 7 & 16.916 & 0.234 & 41 & 19 & 81 & 28 & 0.38 & 0.81 \\
 8 &  16.916 & 0.971 & 25 & 19 & 96 & 22  & 0.45  & 0.57   \\
 9 &  16.920 & 0.963 & 29 & 24 & 72 & 26  & 0.92  & 1.74  \\
10 & 16.924 & 0.241 & 43 & 27 & 91 & 34 & 0.42 & 1.3 \\
11 & 16.924 & 0.265 & 70 & 25 & 86 & 41 & 0.51 & 2.39 \\
12 &  16.924 & 0.948 & 30 & 19 &154 & 24  & 0.90  & 1.39   \\
13 & 16.925 & 0.286 & 41 & 28 & 70 & 34 & 0.4 & 1.27 \\
14 &  16.927 & 0.957 & 36 & 29 &109 & 32  & 4.28  & 11.85  \\
15 & 16.928 & 0.218 & 32 & 26 & 74 & 29 & 0.64 & 1.42 \\
16 &  16.929 & 0.731 & 33 & 21 & 26 & 26  & 0.43  & 0.78   \\ 
17 &  16.934 & 0.777 & 28 & 20 &100 & 23  & 0.55  & 0.82  \\ 
18 &  16.938 & 0.958 & 29 & 24 & 32 & 26  & 0.88  & 1.64 \\
19 &  16.951 & 0.779 & 29 & 24 & 80 & 26  & 1.21  & 2.23 \\
20 &  16.987 & 0.980 & 27 & 22 &124 & 24  & 2.00  & 3.22 \\
21 &  16.994 & 0.931 & 67 & 22 & 88 & 39  & 0.71  & 2.86 \\
22 & 16.996 & 0.417 & 20 & 19 & 131 & 20 & 0.47 & 0.49 \\
23 &  16.998 & 1.001 & 60 & 23 &103 & 37  & 0.68  & 2.54 \\
24 &  17.001 & 0.913 & 41 & 38 & 76 & 39  & 0.64  & 2.70 \\
25 &  17.006 & 0.852 & 25 & 19 & 74 & 22  & 0.45  & 0.58 \\
26 &  17.009 & 0.830 & 33 & 23 & 84 & 28  & 1.90  & 3.99 \\
27 &  17.010 & 0.866 & 28 & 21 & 81 & 24  & 0.41  & 0.65 \\
28 &  17.020 & 1.041 & 34 & 21 &115 & 27  & 0.74  & 1.45 \\
29 &  17.023 & 0.868 & 44 & 32 &  5 & 38  & 0.53  & 2.05 \\
30 &  17.023 & 1.061 & 87 & 34 & 86 & 55  & 0.62  & 5.01 \\
31 & 17.029 & 0.428 & 34 & 26 & 112 & 30 & 0.49 & 1.18 \\
32 &  17.031 & 1.080 & 35 & 24 &104 & 29  & 0.78  & 1.83 \\
33 &  17.031 & 1.066 & 32 & 22 & 69 & 26  & 0.58  & 1.10 \\
34 &  17.032 & 1.042 & 30 & 25 & 79 & 28  & 1.11  & 2.29 \\
35 &  17.035 & 0.750 & 32 & 23 & 87 & 27  & 1.93  & 3.83 \\
36 &  17.038 & 0.874 & 38 & 33 &-50 & 35  & 0.66  & 2.23 \\
37 &  17.063 & 1.095 & 29 & 23 &100 & 26  & 0.51  & 0.93 \\
38 &  17.066 & 0.821 & 60 & 24 & 16 & 38  & 0.54  & 2.13 \\
39 &  17.068 & 0.682 & 39 & 30 & 63 & 34  & 0.84  & 2.67 \\
40 &  17.076 & 1.035 & 50 & 28 & 45 & 37  & 0.74  & 2.78 \\
41 & 17.095 & 0.524 & 24 & 22 & 116 & 23 & 0.53 & 0.77 \\
42 &  17.100 & 0.888 & 34 & 25 &-31 & 29  & 0.44  & 1.00 \\
43 &  17.137 & 0.766 & 30 & 30 & 51 & 30  & 0.78  & 1.93 \\
44 &  17.144 & 0.761 & 33 & 19 & 85 & 25  & 0.47  & 0.81 \\
45 &  17.155 & 0.970 & 32 & 24 & 50 & 28  & 0.66  & 1.38 \\
46 &  17.156 & 1.009 & 35 & 19 &112 & 26  & 0.47  & 0.85 \\
47 &  17.169 & 0.814 & 46 & 32 & 75 & 38  & 0.99  & 3.96 \\
48 &  17.209 & 0.780 & 58 & 23 &141 & 36  & 0.47  & 1.68 \\
49 &  17.216 & 0.832 & 28 & 28 & -2 & 28  & 0.63  & 1.38 \\
50 &  17.217 & 0.821 & 35 & 24 & 93 & 29  & 2.17  & 4.92 \\
51 & 17.222 & 0.396 & 23 & 21 & 126 & 22 & 0.46 & 0.61 \\
\noalign{\smallskip}\hline
\end{tabular}\end{center}
\label{dust-clump-1}
\end{table*}

\begin{table*}
\renewcommand\arraystretch{0.9}
\begin{center}
\tabcolsep 1.2mm\caption{Fitted parameters of $^{12}$CO $J$=1-0, $^{13}$CO $J$=1-0, and C$^{18}$O $J$=1-0 spectra for the selected cores}
\def\temptablewidth{1\textwidth}%
\vspace{-2mm}
\begin{tabular}{llcccccccccr}
\hline\hline\noalign{\smallskip}
ID   &$T_{\rm mb}$(12)   &$FWHM_{\rm spec}$(12)  &$V_{\rm LSR}$(12) & $T_{\rm mb}$(13)   &$FWHM_{\rm spec}$(13)  &$V_{\rm LSR}$(13) & $T_{\rm mb}$(18)   &$FWHM_{\rm spec}$(18)  &$V_{\rm LSR}$(18)\\
  & (K)   & (km s$^{-1}$)   &  (km s$^{-1}$)     &  (K)  & (km s$^{-1}$)  & (km s$^{-1}$) & (K)   & (km s$^{-1}$)   &  (km s$^{-1}$) \\
  \hline\noalign{\smallskip} 
    &&&&&  component 1 &&&&\\    
    \hline\noalign{\smallskip} 
    1 & 12.9(0.4) & 3.0(0.3) & 19.6(0.3) & 6.5(0.2) & 1.6(0.1) & 19.5(0.1) & 1.1(0.2) & 1.2(0.1) & 19.4(0.1) \\
 28 & 23.5(0.4) & 4.1(0.1) & 19.3(0.1) & 13.2(0.1) & 2.5(0.1) & 19.4(0.1) & 3.1(0.1) & 1.9(0.3) & 19.4(0.3) \\
 30 & 19.8(0.4) & 4.2(0.1) & 19.8(0.1) & 12.4(0.2) & 2.5(0.3) & 19.6(0.3) & 3.3(0.2) & 1.7(0.1) & 19.6(0.1) \\
 32 & 20.5(0.4) & 4.0(0.1) & 19.5(0.1) & 11.9(0.2) & 2.6(0.1) & 19.5(0.1) & 2.2(0.2) & 1.9(0.1) & 19.6(0.1) \\
 33 & 20.7(0.4) & 3.9(0.1) & 19.4(0.1) & 12.9(0.2) & 2.5(0.3) & 19.4(0.3) & 3.0(0.2) & 1.8(0.3) & 19.5(0.3) \\
 34 & 24.0(0.5) & 3.9(0.3) & 19.4(0.3) & 13.2(0.1) & 2.3(0.1) & 19.4(0.1) & 2.6(0.1) & 1.9(0.1) & 19.5(0.1) \\
 37 & 19.2(0.6) & 3.9(0.1) & 18.8(0.1) & 11.8(0.2) & 2.4(0.3) & 18.6(0.3) & 2.5(0.2) & 1.6(0.3) & 18.7(0.3) \\
 40 & 26.1(0.3) & 3.0(0.1) & 18.7(0.1) & 12.3(0.1) & 2.1(0.3) & 18.9(0.3) & 1.8(0.1) & 1.8(0.1) & 19.0(0.1) \\
 45 & 24.3(0.5) & 3.0(0.1) & 19.4(0.1) & 10.3(0.2) & 2.1(0.3) & 19.3(0.3) & 1.3(0.2) & 1.8(0.1) & 19.2(0.1) \\
 46 & 19.2(0.4) & 4.3(0.3) & 18.8(0.3) & 6.7(0.2) & 2.5(0.3) & 19.2(0.3) & 0.7(0.1) & 2.2(0.2) & 19.0(0.1) \\
  \hline\noalign{\smallskip} 
    &&&&&  component 2 &&&&\\    
    \hline\noalign{\smallskip} 
  2 & 14.6(0.3) & 3.4(0.3) & 22.8(0.3) & 3.5(0.1) & 4.3(0.3) & 20.4(0.3) & 0.6(0.1) & 3.2(0.3) & 20.2(0.1) \\
  6 & 13.1(0.4) & 4.5(0.3) & 22.2(0.2) & 10.6(0.1) & 3.5(0.0) & 20.6(0.0) & 3.9(0.1) & 2.0(0.1) & 20.3(0.0) \\
  8 & 14.0(0.5) & 4.1(0.1) & 19.7(0.0) & 10.4(0.3) & 3.2(0.3) & 20.5(0.3) & 2.8(0.2) & 2.1(0.3) & 20.7(0.3) \\
  9 & 19.4(0.5) & 4.3(0.3) & 19.4(0.3) & 11.7(0.2) & 3.5(0.3) & 20.6(0.3) & 3.3(0.2) & 2.2(0.3) & 20.7(0.3) \\
  12 & 16.6(0.6) & 3.9(0.3) & 19.5(0.3) & 12.9(0.2) & 3.2(0.0) & 20.4(0.0) & 5.0(0.2) & 2.2(0.3) & 20.5(0.3) \\
  14 & 15.3(0.6) & 3.5(0.3) & 19.5(0.3) & 13.5(0.2) & 3.4(0.3) & 20.5(0.3) & 4.7(0.2) & 2.2(0.0) & 20.5(0.0) \\
 18 & 16.0(0.4) & 3.3(0.1) & 19.9(0.0) & 12.8(0.1) & 2.8(0.0) & 20.5(0.0) & 4.4(0.1) & 1.9(0.0) & 20.6(0.0) \\
 25 & 16.9(0.5) & 3.2(0.3) & 20.4(0.3) & 5.3(0.2) & 2.2(0.3) & 20.6(0.3) & 0.8(0.2) & 1.1(0.3) & 20.6(0.1) \\
 26 & 27.1(0.4) & 4.2(0.2) & 20.0(0.1) & 10.4(0.2) & 3.0(0.3) & 19.8(0.3) & 1.4(0.2) & 2.9(0.2) & 19.8(0.1) \\
 38 & 19.6(0.5) & 4.9(0.1) & 20.6(0.0) & 7.1(0.2) & 3.8(0.3) & 20.9(0.3) & 1.1(0.2) & 1.9(0.3) & 21.1(0.1) \\
 39 & 32.8(0.5) & 3.8(0.3) & 21.8(0.3) & 14.6(0.2) & 2.3(0.0) & 21.6(0.0) & 2.2(0.1) & 1.7(0.3) & 21.6(0.3) \\
 41 & 27.4(0.6) & 3.9(0.1) & 23.9(0.1) & 9.8(0.1) & 2.6(0.0) & 23.4(0.0) & 1.0(0.1) & 1.0(0.2) & 23.4(0.1) \\
 43 & 20.4(0.6) & 3.1(0.0) & 20.4(0.0) & 4.5(0.1) & 2.7(0.3) & 21.0(0.3) & 0.9(0.1) & 2.8(0.3) & 22.6(0.3) \\
 44 & 21.9(0.5) & 3.5(0.1) & 20.5(0.1) & 6.2(0.1) & 2.4(0.3) & 20.8(0.3) & 1.2(0.1) & 1.0(0.3) & 22.9(0.3) \\
 48 & 14.7(0.6) & 3.0(0.3) & 20.3(0.3) & 13.2(0.2) & 2.6(0.0) & 22.7(0.0) & 3.7(0.2) & 1.8(0.3) & 22.8(0.3) \\
 49 & 20.8(0.5) & 3.4(0.1) & 20.2(0.0) & 9.3(0.2) & 3.0(0.3) & 22.8(0.3) & 1.7(0.2) & 1.2(0.2) & 23.2(0.1) \\
 50 & 21.4(0.5) & 3.7(0.3) & 20.6(0.3) & 10.8(0.2) & 3.0(0.1) & 22.8(0.0) & 2.1(0.2) & 2.2(0.2) & 22.8(0.1) \\
 51 & 3.0(0.9) & 2.4(0.6) & 20.3(0.3) & 0.6(0.1) & 1.4(0.3) & 21.5(0.1) & -- & -- & -- \\
  \hline\noalign{\smallskip} 
    &&&&&  component 3 &&&&\\    
    \hline\noalign{\smallskip} 
   3 & 10.3(0.5) & 1.8(0.1) & 26.1(0.0) & 5.5(0.2) & 1.7(0.1) & 25.9(0.0) & -- & -- & -- \\
   4 & 11.0(0.3) & 4.7(0.1) & 22.7(0.0) & 5.3(0.1) & 3.1(0.1) & 23.9(0.0) & -- & -- & -- \\
   5 & 6.6(0.4) & 5.7(0.2) & 24.3(0.1) & 1.5(0.2) & 2.3(0.4) & 23.1(0.2) & -- &-- & -- \\
   7 & 13.3(0.2) & 4.0(0.1) & 22.7(0.0) & 5.3(0.1) & 3.2(0.1) & 24.0(0.0) & -- & -- & -- \\
   10 & 12.8(0.3) & 4.0(0.1) & 22.5(0.0) & 3.7(0.1) & 2.7(0.2) & 24.3(0.1) & -- & -- & -- \\
 11 & 16.2(0.3) & 3.7(0.1) & 22.6(0.0) & 4.9(0.1) & 1.7(0.2) & 23.8(0.2) & -- & -- & -- \\
 13 & 11.9(0.7) & 2.8(1.0) & 21.6(1.0) & 6.2(0.1) & 2.0(0.1) & 24.2(0.0) & -- & -- & -- \\
 15 & 11.3(0.7) & 4.3(0.6) & 21.8(0.4) & 3.7(0.2) & 1.7(0.2) & 24.0(0.2) & -- & -- & -- \\
  16 & 8.9(0.5) & 2.7(0.3) & 25.1(0.3) & 2.2(0.2) & 1.8(0.1) & 25.2(0.1) & -- & -- & -- \\
 17 & 15.3(0.5) & 2.0(0.3) & 29.0(0.3) & 4.8(0.2) & 1.2(0.0) & 28.9(0.0) & -- & -- & -- \\
 19 & 6.9(0.5) & 2.3(0.3) & 25.0(0.3) & 1.6(0.2) & 1.9(0.2) & 25.0(0.1) & -- & -- & -- \\
 20 & 27.3(0.4) & 3.0(0.3) & 24.8(0.3) & 7.8(0.2) & 2.2(0.3) & 24.9(0.3) & 0.7(0.2) & 2.2(0.3) & 25.1(0.2) \\
 21 & 22.1(0.4) & 3.1(0.0) & 26.7(0.0) & 7.8(0.2) & 2.1(0.1) & 26.7(0.0) & 0.9(0.2) & 1.1(0.2) & 26.6(0.1) \\
 22 & 8.7(0.7) & 2.1(0.4) & 24.3(0.1) & 5.0(0.3) & 1.0(0.1) & 25.5(0.0) & -- & -- & -- \\
 23 & 21.9(0.4) & 3.1(0.3) & 25.2(0.3) & 8.9(0.1) & 2.3(0.0) & 25.4(0.0) & 1.2(0.1) & 2.2(0.2) & 25.4(0.1) \\
 24 & 21.6(0.4) & 4.3(0.3) & 26.9(0.3) & 4.9(0.2) & 3.4(0.3) & 26.9(0.3) & -- & -- & --\\
 27 & 11.0(0.4) & 2.1(0.0) & 28.2(0.0) & 4.0(0.3) & 2.0(0.1) & 28.2(0.1) & -- &-- & -- \\
 29 & 24.1(0.4) & 2.2(0.0) & 28.5(0.0) & 7.9(0.1) & 1.7(0.0) & 28.6(0.0) & 1.1(0.2) & 1.4(0.2) & 28.7(0.1) \\
 31 & 12.9(0.6) & 3.6(0.3) & 22.9(0.1) & 4.2(0.1) & 1.2(0.0) & 25.4(0.0) & -- & -- & -- \\
 35 & 17.7(0.3) & 4.4(0.3) & 24.4(0.3) & 4.6(0.1) & 3.4(0.1) & 24.1(0.0) & 0.6(0.1) & 3.0(0.3) & 23.5(0.1) \\
 36 & 16.8(0.5) & 2.1(0.1) & 29.1(0.0) & 6.9(0.2) & 1.7(0.0) & 28.9(0.0) & 0.9(0.1) & 1.4(0.2) & 28.9(0.1) \\
 42 & 25.4(0.4) & 3.7(0.0) & 24.9(0.0) & 11.2(0.2) & 3.1(0.3) & 24.5(0.3) & 2.2(0.2) & 1.8(0.2) & 24.2(0.1) \\
 47 & 22.8(0.4) & 4.4(0.3) & 24.3(0.3) & 10.3(0.1) & 2.1(0.3) & 24.9(0.3) & 1.6(0.1) & 1.8(0.2) & 24.7(0.1) \\
\noalign{\smallskip}\hline
\end{tabular}\end{center}
\label{Table:dust-clump-2}
\end{table*}

\begin{table*}
\renewcommand\arraystretch{0.9}
\begin{center}
\tabcolsep 2.0mm\caption{Calculated parameters of the selected dust cores}
\def\temptablewidth{1\textwidth}%
\vspace{-2mm}
\begin{tabular}{lcccccccccccccr}
\hline\hline\noalign{\smallskip}
ID    & $R_{\rm eff}$ & Mass & $n_{\rm_{H_{2}}}$ &$T_{\rm ex}$  &$\tau$($^{13}$CO)   & $\sigma_{\rm _{Therm}}$ & $\sigma_{\rm_{NT}}$ & $\sigma_{\rm_{v}}$  & Type \\
    & (pc)   & (M$_{\odot}$) 
& (10$^{4}$ cm$^{-3}$) & (K)   &   & (km s$^{-1}$)  & (km s$^{-1}$)  & (km s$^{-1}$)    & \\
  \hline\noalign{\smallskip} 
    &&&&&  component 1 &&&&\\    
    \hline\noalign{\smallskip} 
   1 & 0.10 &  20.9(1.0) & 8.6(0.4) & 17.0(0.5) & 0.7(0.1) & 0.22(0.01) & 0.51(0.06) & 0.55(0.03)  & \hii/PDR \\
 28  & 0.12 &  26.2(1.2) & 5.1(0.2) & 28.1(0.5) & 0.8(0.1) & 0.29(0.01) & 0.80(0.14) & 0.85(0.07)  & quiescent\\
 30  & 0.25 & 110.8(5.9) & 2.5(0.1) & 24.2(0.5) & 0.9(0.1) & 0.27(0.01) & 0.73(0.01) & 0.78(0.01)  & quiescent \\
 32  & 0.13 &  38.8(2.0) & 6.1(0.3) & 24.9(0.5) & 0.8(0.1) & 0.27(0.01) & 0.81(0.05) & 0.86(0.02)  & quiescent \\
 33 & 0.12 &  23.1(1.1) & 5.0(0.2) & 25.1(0.5) & 0.9(0.1) & 0.27(0.01) & 0.78(0.14) & 0.82(0.07)  & quiescent \\
 34 & 0.13 &  40.4(2.4) & 7.0(0.4) & 28.6(0.6) & 0.7(0.1) & 0.29(0.01) & 0.78(0.03) & 0.83(0.02)  & quiescent \\
 37  & 0.12 &  21.1(1.6) & 4.6(0.3) & 23.6(0.7) & 0.9(0.1) & 0.26(0.01) & 0.67(0.14) & 0.72(0.07)  & quiescent \\
 40  & 0.17 &  44.1(1.5) & 3.4(0.1) & 30.8(0.3) & 0.6(0.1) & 0.30(0.01) & 0.77(0.03) & 0.82(0.02)  & quiescent \\
 45  & 0.13 &  24.0(1.3) & 4.2(0.2) & 28.9(0.5) & 0.5(0.1) & 0.29(0.01) & 0.74(0.03) & 0.80(0.02)  & \hii/PDR \\
 46  & 0.12 &  19.3(0.9) & 4.2(0.2) & 23.6(0.5) & 0.4(0.1) & 0.26(0.01) & 0.94(0.10) & 0.98(0.05)  & \hii/PDR\\
  \hline\noalign{\smallskip} 
    &&&&&  component 2 &&&&\\    
    \hline\noalign{\smallskip} 
   2 &  0.11 &  49.8(2.1) & 13.7(0.6) & 18.7(0.4) & 0.3(0.1)  & 0.24(0.01) & 1.34(0.11) & 1.36(0.06)  & \hii/PDR \\
  6  &  0.13  &  51.5(2.7) & 8.0(0.4) & 17.2(0.5) & 1.5(0.1)  & 0.23(0.01) & 0.84(0.03) & 0.87(0.02)  & quiescent \\
  8  &  0.10  &  19.2(1.1) & 6.9(0.4) & 18.1(0.6) & 1.2(0.1)  & 0.23(0.01) & 0.89(0.14) & 0.92(0.07)  & quiescent \\
  9  &  0.12  &  39.4(2.4) & 8.5(0.5) & 23.8(0.6) & 0.9(0.1)  & 0.26(0.01) & 0.95(0.14) & 0.99(0.07)   & \hii/PDR \\
  12  &  0.11  &  38.0(2.7) & 10.4(0.7) & 20.8(0.7) & 1.3(0.1)  & 0.25(0.01) & 0.91(0.14) & 0.95(0.07)  & quiescent \\
  14  &  0.14  & 357.3(25.5) & 41.5(3.0) & 19.5(0.7) & 1.8(0.1)  & 0.24(0.01) & 0.95(0.02) & 0.98(0.01)  & \hii/PDR \\
 18  &  0.12  &  46.8(2.3) & 10.1(0.5) & 20.2(0.5) & 1.4(0.1)  & 0.24(0.01) & 0.81(0.01) & 0.85(0.01)  & \hii/PDR \\
 25  &  0.10  &  15.5(1.0) & 5.5(0.4) & 21.2(0.7) & 0.4(0.1)  & 0.25(0.01) & 0.45(0.11) & 0.52(0.06)  & \hii/PDR \\
 26  &  0.13  &  61.3(3.0) & 10.6(0.5) & 31.8(0.5) & 0.5(0.1)  & 0.31(0.01) & 1.22(0.10) & 1.25(0.05)  & \hii/PDR \\
 38  &  0.17  &  47.7(2.9) & 3.3(0.2) & 24.0(0.6) & 0.4(0.1)  & 0.27(0.01) & 0.82(0.13) & 0.86(0.07)  & quiescent \\
 39  &  0.15  &  33.4(2.0) & 3.2(0.2) & 37.8(0.6) & 0.6(0.1)  & 0.33(0.01) & 0.72(0.14) & 0.79(0.07)  & \hii/PDR\\
 41  & 0.11  &  11.9(0.8) & 3.6(0.2) & 32.2(0.7)  & 0.4(0.1)  & 0.31(0.00) & 0.43(0.08) & 0.53(0.04)  & protostellar \\
 43  &  0.13  &  41.2(2.9) & 5.8(0.4) & 24.8(0.7) & 0.2(0.1)  & 0.27(0.01) & 1.19(0.14) & 1.22(0.07)  & \hii/PDR \\
 44  &  0.11  &  15.9(0.9) & 3.9(0.2) & 26.4(0.6) & 0.3(0.1)  & 0.28(0.01) & 0.43(0.14) & 0.52(0.07)  &\hii/PDR \\
 48  &  0.16  &  53.0(3.9) & 4.3(0.3) & 18.9(0.7) & 1.9(0.1)  & 0.24(0.01) & 0.78(0.14) & 0.81(0.07)  & quiescent \\
 49  &  0.13  &  28.8(1.7) & 5.0(0.3) & 25.2(0.6) & 0.6(0.1)  & 0.27(0.01) & 0.49(0.07) & 0.56(0.03) ) & quiescent \\
 50  &  0.13  &  99.3(5.6) & 15.5(0.9) & 25.9(0.6) & 0.7(0.1)  & 0.28(0.01) & 0.92(0.07) & 0.96(0.04)  & \hii/PDR \\
 51  & 0.10  & 75.1(19.8) & 26.0(6.9) & 9.0(2.6) & 0.1(0.1) & -- & -- & --  & \hii/PDR \\
  \hline\noalign{\smallskip} 
    &&&&&  component 3 &&&&\\    
    \hline\noalign{\smallskip} 
  
 3  & 0.12  &  52.8(3.7) & 11.1(0.8) & 14.2(0.7) & 0.7(0.1) & 0.20(0.01) & 0.34(0.07) & 0.40(0.03)  & quiescent \\
 4  & 0.17  &  54.7(2.4) & 3.7(0.2) & 18.1(0.4) & 0.6(0.1) & -- & -- & --  & quiescent \\
 5  &  0.11  &  97.5(6.6) & 23.4(1.6) & 10.3(0.7) & 0.2(0.1) & -- & -- & -- & \hii/PDR \\
 7  & 0.13  &  23.1(0.8) & 3.9(0.1) & 20.5(0.4) & 0.5(0.1) & -- & -- & --  & quiescent  \\
 10  & 0.15  &  38.5(1.7) & 3.6(0.2) & 20.0(0.4) & 0.3(0.1) & -- & -- & --   & quiescent\\
 11  & 0.19  &  56.0(2.1) & 3.0(0.1) & 23.6(0.4) & 0.3(0.1) & -- & -- & --   & quiescent \\
 13  & 0.15  &  40.4(4.4) & 3.8(0.4) & 19.0(1.1) & 0.7(0.1) & -- & -- & --   & \hii/PDR \\
 15  & 0.13  &  47.6(5.6) & 7.2(0.8) & 18.4(1.2) & 0.4(0.1) & -- & -- & --  & protostellar \\
 16 &  0.12  &  47.5(3.3) & 10.1(0.7) & 12.7(0.7) & 0.3(0.1) & -- & -- & --  & \hii/PDR \\
 17  &  0.10  &  25.3(1.8) & 7.7(0.5) & 19.5(0.7) & 0.4(0.1) & -- & -- & -- & \hii/PDR \\
 19  &  0.12  & 192.5(15.9) & 41.0(3.4) & 10.6(0.8) & 0.2(0.1) & -- & -- & --  & \hii/PDR \\
 20  &  0.11  &  49.0(2.1) & 13.5(0.6) & 32.0(0.4) & 0.3(0.1) &  0.31(0.00) & 0.94(0.15) & 0.99(0.07)  & \hii/PDR  \\
 21  &  0.18 &  56.2(2.8) & 3.5(0.2) & 26.6(0.5) & 0.4(0.1) &  0.28(0.00) & 0.48(0.10) & 0.55(0.05)  & \hii/PDR \\
 22  & 0.09  &  21.3(2.6) & 9.8(1.2) & 15.6(1.2) & 0.8(0.1) & -- & -- & --   & quiescent \\
 23  &  0.17  &  50.4(2.2) & 3.7(0.2) & 26.4(0.4) & 0.5(0.1) & 0.28(0.00) & 0.95(0.07) & 0.99(0.04)  & \hii/PDR \\
 24  &  0.18  &  55.1(2.9) & 3.4(0.2) & 26.1(0.5) & 0.2(0.1) & -- & -- & -- & \hii/PDR\\
 27  &  0.11  &  30.3(1.5) & 8.1(0.4) & 14.9(0.5) & 0.4(0.1) & -- & -- & -- & \hii/PDR \\
 29  &  0.17  &  36.8(1.6) & 2.5(0.1) & 28.7(0.4) & 0.4(0.1) &  0.29(0.00) & 0.57(0.09) & 0.64(0.04)  & \hii/PDR \\
 31  & 0.14  &  34.7(3.5) & 4.7(0.5) & 20.1(1.0) & 0.4(0.1) & -- & -- & --  & quiescent \\
 35  &  0.12  &  97.8(3.5) & 18.6(0.7) & 22.0(0.4) & 0.3(0.1)  & 0.25(0.00) & 1.26(0.12) & 1.28(0.06) & \hii/PDR \\
 36  &  0.16  &  61.3(4.0) & 5.3(0.3) & 21.0(0.6) & 0.5(0.1)  & 0.25(0.00) & 0.61(0.09) & 0.66(0.04) & \hii/PDR \\
 42  &  0.13  &  16.9(0.8) & 2.6(0.1) & 30.1(0.5) & 0.5(0.1)  & 0.30(0.00) & 0.78(0.07) & 0.83(0.03)  & \hii/PDR \\
 47  &  0.17  &  75.8(3.6) & 5.1(0.2) & 27.3(0.5) & 0.6(0.1)  &  0.28(0.00) & 0.75(0.09) & 0.80(0.05)  & protostellar \\
\noalign{\smallskip}\hline
\end{tabular}\end{center}
\label{Table:dust-clump-3}
\end{table*}

\end{document}